\documentclass[12pt, a4paper]{article}
\pdfoutput=1
\usepackage[left=20mm,right=20mm,top=20mm,bottom=25mm]{geometry}
\usepackage[affil-it]{authblk}
\usepackage{hyperref}
\usepackage{amsmath}
\usepackage{amssymb}
\usepackage{graphicx}
\usepackage{cite}
\usepackage{tikz}
\usepackage{microtype}
\usepackage{subcaption}

\usetikzlibrary{arrows}
\usetikzlibrary{decorations.pathmorphing}
\usetikzlibrary{decorations.markings}
\usetikzlibrary{calc}

\tikzset{
  >=stealth',
  midarrow/.style={
    postaction={
      decorate,
      decoration={markings, mark=at position .55 with {\arrow{>}}}
    }
  },
  fermion/.style=midarrow,
  photon/.style={
    decorate,
    decoration={snake, amplitude=2pt, segment length=8pt}
  },
  boson/.style={
    decorate,
    decoration={snake, amplitude=2pt, segment length=8pt}
  },
  gluon/.style={
    decorate,
    decoration={coil, amplitude=4pt, segment length=5pt}
  },
  scalar/.style=densely dashed,
  arrowsnake/.style={
    preaction={photon, draw},
    postaction=midarrow
  }
}

\hypersetup{
  colorlinks,
  linkcolor={blue!50!black},
  citecolor={blue!50!black},
  urlcolor={blue!50!black}
}

\newcommand{\ch}{\tilde \chi_1}
\newcommand{\abs}[1]{\lvert #1 \rvert}
\newcommand{\nuc}[2]{$^{#2}$#1}

\begin{document}

\title{Quasistable charginos in ultraperipheral proton-proton collisions at the
LHC}

\author[1,2]{S.~I.~Godunov~\thanks{sgodunov@itep.ru}}
\author[1,3,4]{V.~A.~Novikov~\thanks{novikov@itep.ru}}
\author[1,5]{A.~N.~Rozanov~\thanks{rozanov@cppm.in2p3.fr}}
\author[1,3,4]{M.~I.~Vysotsky~\thanks{vysotsky@itep.ru}}
\author[1,6]{E.~V.~Zhemchugov~\thanks{zhemchugov@itep.ru}}

\affil[1]{\small Institute for Theoretical and Experimental Physics, 117218,
Moscow, Russia}
\affil[2]{\small Novosibirsk State University, 630090, Novosibirsk, Russia}
\affil[3]{\small National Research University Higher School of Economics,
101978, Moscow, Russia}
\affil[4]{\small Moscow Institute of Physics and Technology, 141701,
Dolgoprudny, Moscow region, Russia}
\affil[5]{\small Centre de Physique des Particules de Marseille, CPPM,
Aix-Marseille Universite, CNRS/IN2P3, Marseille, France}
\affil[6]{\small Moscow Engineering Physics Institute, 115409, Moscow, Russia}

\date{}

\maketitle

\vspace*{-1cm}

\begin{abstract}
We propose a model-independent approach for the search of charged long-lived
particles produced in ultraperipheral collisions at the LHC. The main idea is to
improve event reconstruction at ATLAS and CMS with the help of their forward
detectors.  Detection of both scattered protons in forward detectors allows
complete recovery of event kinematics. Though this requirement reduces the
number of events, it greatly suppresses the background, including the large
background from the pile-up.
\end{abstract}

\section{Introduction}

The Large Hadron Collider (LHC) can be considered as a photon-photon collider
with the photons produced in ultraperipheral collisions (UPC) of charged
particles: protons or heavy ions. In such collisions the colliding particles
pass near each other exchanging photons; the particles remain intact after the
collision.  Enormous collision energy achieved at the LHC permits treatment of
the particles' electromagnetic fields as bunches of real photons distributed
according to a well-known spectrum. This approximation is known as the
equivalent photon approximation~(EPA)~\cite{zeit.phys.29.315, zeit.phys.88.612,
williams, pzs6.244} (see also~\cite{jetp11-388, rev.mod.phys.4.615, pepan4.239,
prep15.181}).

Ultraperipheral collisions are a promising source of New Physics events for the
kinds of physics that can appear in photon fusion. They feature clear
experimental signature with only the photon fusion result and the two initial
particles in the final state. The colliding particles scatter at a very small
angle and escape the detector through the beam pipe. They can be registered with
forward detectors---ATLAS Forward Proton Detector
(AFP)~\cite{atlas-tdr-024-2015} or CMS-TOTEM Precision Proton
Spectrometer~\cite{totem-tdr-003}. These detectors are located at the distance
of $\approx 200$~m from the interaction point along the beam pipe, and they can
be moved as close as a few millimeters from the beam.  Forward detectors can
detect a proton with efficiency near 100\% if its fractional momentum loss, $\xi
\equiv \Delta p / p$, is in the range $0.015 < \xi <
0.15$~\cite{atlas-tdr-024-2015, totem-tdr-003}. In the original FP420
proposal~\cite{jinst4-t10001} forward detectors were placed at 420~m from the
interaction point, and the corresponding fractional momentum loss range was at
smaller values $0.002 < \xi < 0.020$. This position at 420~m was not retained in
the actual AFP detector, but can be used for estimations of sensitivities. The
corresponding energy losses are presented in Table~\ref{t:energy-losses}.
Unfortunately, a heavy ion from lead-lead collisions with the energy of
$5.02$~TeV/nucleon pair cannot be detected in forward detectors because the
production cross section and EPA spectrum are highly suppressed at $\xi \gtrsim
0.002$.
\begin{table}
  \centering
  \caption{Energy losses required for a particle to be detected in the forward
  detector placed at different distances from the interaction point (IP).}
  \begin{tabular}{lr@{--}lr@{--}l}
    \hline \hline
    Distance from the IP, m & \multicolumn{2}{c}{200} & \multicolumn{2}{c}{420} \\
    $\xi$ range & $0.015$ & $0.15$ & $0.002$ & $0.02$ \\
    $6.5$~TeV $p$ energy loss, GeV & $97.5$ & 975 & 13 & 130 \\
    $0.5$~PeV \nuc{Pb}{208} energy loss, TeV & $7.8$ & 78 & $1.0$ & 10 \\
    \hline \hline
  \end{tabular}
  \label{t:energy-losses}
\end{table}

Photon flux in UPC is proportional to $(Z_1 e)^2 (Z_2 e)^2$, where $Z_1 e$ and
$Z_2 e$ are electric charges of the colliding particles.  In this respect,
collisions of heavy ions, e.g. lead ions with $Z = 82$, look much more promising
for the search of New Physics. However, in order for the process of photon
emission to be coherent, photon virtuality $q^2$ where $q$ is the photon
4-momentum has to be smaller than square of the inverse of the charge radius. In
the case of proton, calculation based on its electromagnetic form factor results
in $\hat q = 0.20$~GeV~\cite{1806.07238}, where $\hat q$ is the maximum momentum
of a virtual photon in the proton rest frame. In the laboratory reference frame
the maximum photon energy is $\hat q \gamma$ where $\gamma$ is the Lorentz
$\gamma$-factor of the proton; for a $6.5$~TeV proton $\hat q \gamma = 1.4$~TeV.
For the lead ion, $\hat q = 30$~MeV~\cite{1806.07238}, so in lead-lead
collisions with the energy $5.02~\text{TeV}/\text{nucleon pair}$, maximum photon
energy is $80$~GeV.  Photons with higher energy are produced as well, but their
production is suppressed by the nucleus form factor, thus greatly reducing the
benefits from higher photon flux in a collision of heavy ions.  Nevertheless,
the production cross section for a system with invariant mass about 100~GeV is
several orders of magnitude larger in lead-lead collisions than in proton-proton
collisions~\cite{1806.07238}.

A good example of New Physics that can be searched in UPC is supersymmetry
(SUSY)~\cite{hep-ph-9402302, 0806.1097, 1002.2857, 1110.4320, 1811.06465}. The
supersymmetric partners of the electroweak bosons are six particles: four
neutralinos and two charginos. Let $\ch^0$ be the lightest neutralino and
$\ch^\pm$ be the lightest chargino. At present, chargino and neutralino with
masses below $\sim 1$~TeV are excluded in a large region of SUSY parameters by
the LHC results~\cite[\textsection{}110.5]{pdg}.  However, most of the searches
are much less sensitive to the case when the masses of the lightest chargino and
the lightest neutralino are approximately equal. In particular, in the framework
of the MSSM, when $m_{\ch^\pm} - m_{\ch^0} \lesssim 2$~GeV, the bound $m_{\tilde
\chi_{1}^{\phantom{.}}} > 92$~GeV comes from the LEP
experiments~\cite{LEPSUSYWG/02-04.1}.  At this mass scale, the possibility that
$m_{\ch^\pm} < m_{\ch^0}$ is excluded, since then the chargino would be stable
(assuming $R$-parity conservation), and the charginos remaining after the Big
Bang and/or produced in cosmic rays would form hydrogen-like atoms that would be
observed in sea water~\cite{npb206-333, prd41-2074, prl68-1116, prd47-1231} (see
also~\cite{hep-ph/0202252}).\footnote{Concentration of relic charginos would be
the same as neutralinos in the standard scenario (where neutralino is the LSP).
For chargino mass of the order of 100~GeV, this value would be of the same order
of magnitude as protons concentration, and it is in dramatic contradiction with,
e.g., the bound of $10^{-28}$ times protons concentration from
Ref.~\cite{npb206-333}.} Thus, in what follows we consider only the case when
the lightest chargino is (slightly) heavier than the lightest neutralino.  Such
compressed chargino-neutralino spectrum is realized in the following two cases:
$M_2 \ll M_1, \mu$ (wino-like) or $\mu \ll M_1, M_2$ (higgsino-like), where
$M_1$ is the bino mass parameter, $M_2$ is the wino mass parameter, and $\mu$ is
the higgsino mass parameter.

In this scenario, chargino might live long enough to fly through the detector
and decay outside if they are produced at the LHC. Such particles are called
long-lived charged particles (LLCP).  In this paper we suggest an approach for
the search of LLCP using forward detectors of the ATLAS and CMS collaborations.
Although LLCP appear in a variety of models of New Physics, we find SUSY with
compressed mass scenario to be the most interesting. Nevertheless, our results
can be applied to LLCP of any nature.

There are many searches for long-lived particles in inelastic processes at
the LHC~\cite{1101.1645, 1106.4495, 1205.0272, 1305.0491, 1411.6795, 1502.02522,
1506.05332, 1506.09173, 1604.04520, 1609.08382, 1808.04095, 1902.01636}. The
unobservation of LLCP at the LHC allows the experiments to set model-independent
constraints on fiducial LLCP production cross sections. These constraints are
then reinterpreted in a particular model to derive bounds on LLCP integrated
production cross sections and masses.  Therefore, the bounds established in
these papers strongly depend on the choice of the supersymmetric model, e.g., on
squark masses and\slash{}or chargino coupling to $Z$. Further extension of the
model with New Physics such as extra Higgs fields or $Z'$ bosons will affect
these bounds as well. In ultraperipheral collisions, chargino production is
mediated by photons which couple to chargino in a model independent way.
Consequently, UPC provide us with a way to set model-independent bounds on the
masses of charginos (or other long-lived charged particles).

Small cross sections of UPC processes prevent observation of LLCP in the
previous searches~\cite{1101.1645, 1106.4495, 1205.0272, 1305.0491, 1411.6795,
1502.02522, 1506.05332, 1506.09173, 1604.04520, 1609.08382, 1808.04095,
1902.01636}. See Appendix~\ref{s:llp-searches} for the detailed discussion.

The region of SUSY parameters with $m_{\ch^0} \approx m_{\ch^\pm} \equiv m_\chi
\sim 100$~GeV can be probed at the LHC in UPC of both protons and heavy ions.
Let us consider the cross sections (Section~\ref{s:production}), the search
strategy (Section~\ref{s:search}), the background (Section~\ref{s:background})
for chargino production, and the accessible chargino masses and
lifetimes~(Section \ref{s:lifetimes}). In Appendix~\ref{s:llp-searches} we
discuss why papers~\cite{1101.1645, 1106.4495, 1205.0272, 1305.0491, 1411.6795,
1502.02522, 1506.05332, 1506.09173, 1604.04520, 1609.08382, 1808.04095,
1902.01636} do not exclude the production of charged long-lived particles with
masses 100--200~GeV in ultraperipheral collisions.

\section{Production cross section}

\label{s:production}

\begin{figure}[!tbhp]
  \centering
  \begin{tikzpicture}

    \coordinate (A)   at (-0.7,  1.2);
    \coordinate (B)   at (   0,  0.5);
    \coordinate (C)   at (   0, -0.5);
    \coordinate (D)   at (-0.7, -1.2);
    \coordinate (P1i) at (-1.2,  1.2);
    \coordinate (P1o) at (   1,  1.2);
    \coordinate (P2i) at (-1.2, -1.2);
    \coordinate (P2o) at (   1, -1.2);
    \coordinate (M1)  at (   1,  0.5);
    \coordinate (M2)  at (   1, -0.5);

    \draw[fermion] (P1i) node [left] {$p$} -- (A);
    \draw[fermion] (A)   -- (P1o) node [right] {$p$};
    \draw[fermion] (P2i) node [left] {$p$} -- (D);
    \draw[fermion] (D)   -- (P2o) node [right] {$p$};
    \draw[photon]  (A)   -- (B);
    \draw[photon]  (D)   -- (C);
    \draw[fermion] (M1)  node [right] {$\ch^+$} -- (B);
    \draw[fermion] (B)   -- (C);
    \draw[fermion] (C)   -- (M2) node [right] {$\ch^-$};
  \end{tikzpicture}
  ~
  \begin{tikzpicture}
    \coordinate (A)   at (-0.7,  1.2);
    \coordinate (B)   at (   0,  0.5);
    \coordinate (C)   at (   0, -0.5);
    \coordinate (D)   at (-0.7, -1.2);
    \coordinate (P1i) at (-1.2,  1.2);
    \coordinate (P1o) at (   1,  1.2);
    \coordinate (P2i) at (-1.2, -1.2);
    \coordinate (P2o) at (   1, -1.2);
    \coordinate (M1)  at (   1,  0.5);
    \coordinate (M2)  at (   1, -0.5);

    \draw[fermion] (P1i) node [left] {$p$} -- (A);
    \draw[fermion] (A)   -- (P1o) node [right] {$p$};
    \draw[fermion] (P2i) node [left] {$p$} -- (D);
    \draw[fermion] (D)   -- (P2o) node [right] {$p$};
    \draw[photon]  (A)   -- (C);
    \draw[photon]  (D)   -- (B);
    \draw[fermion] (M1)  node [right] {$\ch^+$} -- (B);
    \draw[fermion] (B)   -- (C);
    \draw[fermion] (C)   -- (M2) node [right] {$\ch^-$};
  \end{tikzpicture}  
  \caption{Leading order Feynman diagrams for chargino production in an
  ultraperipheral collision of two protons.}
  \label{f:collision}
\end{figure}
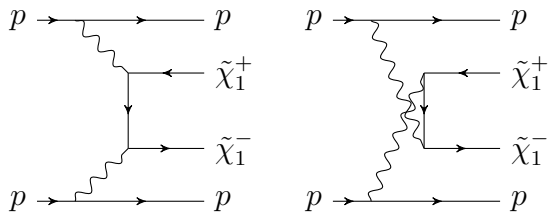

Consider production of a pair of charginos in an UPC of two identical particles
with charge $Ze$. The leading order Feynman diagrams of this process for protons
are presented in Fig.~\ref{f:collision}. Collision is mediated by approximately
real photons emitted from the colliding particles.  The equivalent photon
approximation provides the momentum distribution of these
photons~\cite{landau-lifshitz}:
\begin{equation}
  n(\vec q_\perp, \omega) \, \mathrm{d}^2 q_\perp \, \mathrm{d} \omega
  = \frac{Z^2 \alpha}{\pi^2}
    \frac{\vec q_\perp^{\; 2}}
         {\omega \left( \vec q_\perp^{\; 2} + \dfrac{\omega^2}{\gamma^2}
         \right)^2}
    \left\lvert
      F \left( \vec q_\perp^{\; 2} + \frac{\omega^2}{\gamma^2} \right)
    \right\rvert^2
    \, \mathrm{d}^2 q_\perp \, \mathrm{d} \omega,
 \label{epa-distribution}
\end{equation}
where $\omega$ is the photon energy in the laboratory frame, $\vec q_\perp$ is
the transverse component of the photon momentum, $\gamma$ is the Lorentz
factor of the source particle, $F$ is the form factor originating from the
vertex involving the particle which emit photons. Let us note that $\vec
q_\perp^{\; 2} + (\omega / \gamma)^2 = -q^2$, where $q$ is the photon
4-momentum.

When discussing the form factors it is convenient to use photon 3-momentum in
the rest frame of the source particle $\vec q = (\vec q_\perp, \omega /
\gamma)$. For the proton, the Dirac form factor is~\cite{prep550-1}
\begin{equation}
  F(\vec q^{\; 2})
  = G_D(\vec q^{\; 2}) \left[ 1 + \frac{(\mu_p - 1) \tau}{1 + \tau} \right],
\end{equation}
where
\begin{equation}
  G_D(\vec q^{\; 2}) \equiv \frac{1}{(1 + \vec q^{\; 2} / \Lambda^2)^2}
\end{equation}
is the dipole form factor, $\mu_p = 2.79$ is the proton magnetic moment, $\tau =
\vec q^{\; 2} / 4 m_p^2$, $m_p$ is the proton mass, and $\Lambda^2 =
0.71~\text{GeV}^2$.  Since in an UPC $\abs{q^2} \lesssim \Lambda_\text{QCD}^2
\ll 4 m_p^2$, the magnetic form factor contribution can be neglected. In this
case $F(\vec q^{\; 2}) \approx G_D(\vec q^{\; 2})$, and the equivalent photon
spectrum is given by~\cite{1806.07238}
\begin{equation}
  n_p(\omega) \, \mathrm{d} \omega
  = \frac{\alpha}{\pi}
    \left[
      (4a + 1) \ln \left( 1 + \frac{1}{a} \right)
      - \frac{24 a^2 + 42 a + 17}{6 (a + 1)^2}
    \right]
    \frac{\mathrm{d} \omega}{\omega},
  \label{spectrum}
\end{equation}
where $a = (\omega / \Lambda \gamma)^2$.

Heavy nucleus form factor is more complicated. The most accurate description of
nucleus charge distribution appears to be in the form of Bessel
decomposition~\cite{npa235-219}:
\begin{equation}
  \rho(r) = \sum\limits_{n=1}^N a_n \, j_0(n \pi r / R) \, \theta(R - r),
\end{equation}
where $j_0(x) = \sin x / x$ is the spherical Bessel function of order zero,
$\theta(x)$ is the Heaviside step function, $a_n$ and $R$ are parameters of the
decomposition. The form factor is the Fourier transform of the charge
distribution:
\begin{equation}
  F(\vec q^{\; 2})
  = \frac{\int \rho(r) \mathrm{e}^{i \vec q \vec r} \mathrm{d}^3 r}
         {\int \rho(r) \mathrm{d}^3 r}
  = \frac{\sin \abs{\vec q \,} R}{\abs{\vec q \,} R}
    \cdot
    \frac{\sum\limits_{n=1}^N \frac{(-1)^n a_n}{n^2 \pi^2 - \vec q^{\; 2} R^2}}
         {\sum\limits_{n=1}^N \frac{(-1)^n a_n}{n^2 \pi^2}}.
\end{equation}
Numerical values of $a_n$ and $R$ are provided in Ref.~\cite{adndt36-495}. The
corresponding equivalent photon spectrum $n_\text{Pb}(\omega)$ is calculated
through numerical integration of Eq.~\eqref{epa-distribution}.

\begin{figure}[!t]
  \centering
  \includegraphics{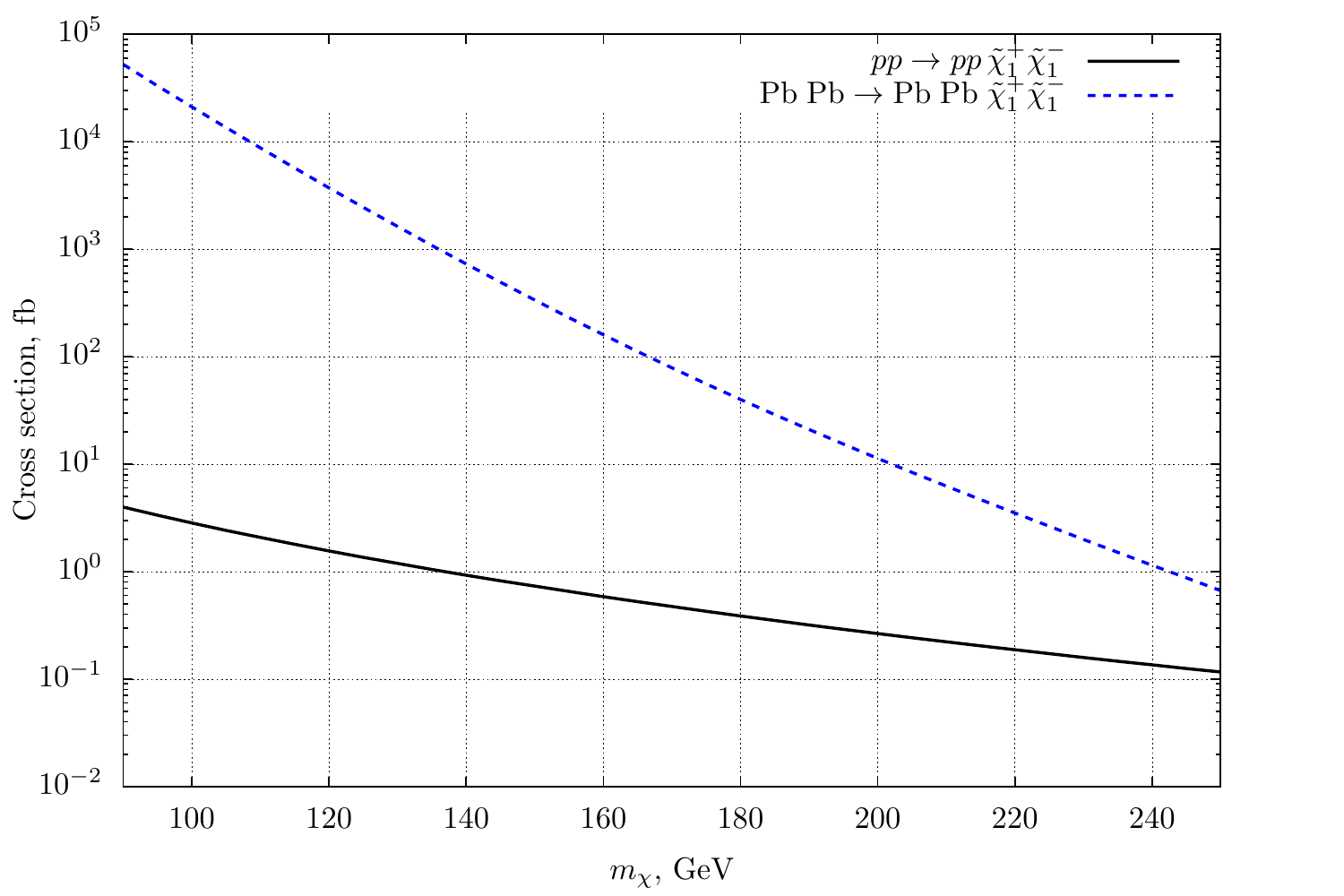}
  \caption{
    Cross sections for chargino pair production in proton-proton UPC with the
    collision energy 13~TeV and lead-lead UPC with the collision energy
    5.02~TeV/nucleon pair.
  }
  \label{f:xsection-m}
\end{figure}

Production of charginos in photon fusion is described by the Breit-Wheeler cross
section~\cite{pr46-1087},
\begin{equation}
  \sigma(\gamma \gamma \to \ch^+ \ch^-)
  = \frac{4 \pi \alpha^2}{s}
    \left[
      \left( 1 + \frac{4 m_\chi^2}{s} - \frac{8 m_\chi^4}{s^2} \right)
      \ln \frac{1 + \sqrt{1 - 4 m_\chi^2 / s}}
               {1 - \sqrt{1 - 4 m_\chi^2 / s}}
      - \left( 1 + \frac{4 m_\chi^2}{s} \right)
        \sqrt{1 - \frac{4 m_\chi^2}{s}}
    \right],
\end{equation}
where $\sqrt{s} \equiv \sqrt{4 \omega_1 \omega_2}$ is the invariant mass of the
pair of charginos, $\omega_1$ and $\omega_2$ are photons energies. Cross section
for charginos production in ultraperipheral collisions is
\begin{equation}
  \sigma(N N \to N N \ch^+ \ch^-)
  = \int\limits_0^\infty \int\limits_0^\infty
      \sigma(\gamma \gamma \to \ch^+ \ch^-) \,
      n_N(\omega_1) \, n_N(\omega_2)
      \, \mathrm{d} \omega_1 \, \mathrm{d} \omega_2,
  \label{epa-xsection}
\end{equation}
where $N$ is the colliding particle, $n_N(\omega)$ is its equivalent photon
spectrum. For $m_\chi = 100$~GeV,
\begin{equation}
  \sigma(p p \to p p \ch^+ \ch^-) = 2.84~\text{fb},\ 
  \sigma(\text{Pb} \; \text{Pb} \to \text{Pb} \; \text{Pb} \; \ch^+ \ch^-)
  = 21.2~\text{pb},\footnote{
     Finite proton radius suppresses this cross section by 10--20\% due to the
     so-called survival factor~\cite[Table~1]{1410.2983},~\cite{1508.02718}. For
     lead ions the suppression should be larger~\cite{prd42-3690, 1810.06567}.
     In this paper we are interested in feasibility of our approach to the
     search of new charged particles.  Accurate calculations with the survival
     factor taken into account is the subject of a subsequent paper.
  }
  \label{epa-xsection:values}
\end{equation}
where the proton-proton collision energy is 13~TeV, and lead-lead collision
energy is $5.02$~TeV/nuc\-leon pair (these parameters correspond to the currently
available LHC data). Cross section dependence on chargino mass is presented in
Fig.~\ref{f:xsection-m}. At higher masses chargino pair production in lead-lead
collisions is heavily suppressed by the lead ion form factor.

In order for both colliding particles to be detected in forward detectors (FD),
their momentum loss $\xi = \Delta p / p$ has to be in the interval
$\xi_\text{min} < \xi < \xi_\text{max}$, where $\xi_\text{min} = 0.015$ and
$\xi_\text{max} = 0.15$ for the ATLAS and CMS
experiments~\cite{atlas-tdr-024-2015, totem-tdr-003} (see
Table~\ref{t:energy-losses}). The corresponding cross section is given by
formula~\eqref{epa-xsection} with cuts on photon energies:
\begin{equation}
  \sigma_\text{FD}(N N \to N N \ch^+ \ch^-)
  = \int\limits_{\xi_\text{min} E}^{\xi_\text{max} E}
      \int\limits_{\xi_\text{min} E}^{\xi_\text{max} E}
        \sigma(\gamma \gamma \to \ch^+ \ch^-) \,
        n_N(\omega_1) \, n_N(\omega_2)
        \, \mathrm{d} \omega_1 \, \mathrm{d} \omega_2,
\end{equation}
where $2 E$ is the collision energy. For the same parameters as
in~\eqref{epa-xsection:values},
\begin{equation}
  \sigma_\text{FD}(pp \to pp \ch^+ \ch^-) = 0.80~\text{fb}.
  \label{epa-xsection:fd}
\end{equation}

For lead ions, according to Eq.~\eqref{epa-xsection:values}, with the current
integrated luminosity $2.5~\text{nb}^{-1}$~%
\cite{atlas-luminosity, cms-luminosity}, there will be $0.053$ events. To
observe chargino in lead-lead collisions, the integrated luminosity has to be
tremendously increased. If the luminosity could be increased by three orders of
magnitude, there would be about 50 events. For a lead ion to survive in an UPC,
its energy loss should not be much greater than $\approx
100$~GeV~\cite{1806.07238}. The corresponding value of $\xi$ is $1.9 \cdot
10^{-4}$. It makes detection of a lead ion in a forward detector impossible (see
Table~\ref{t:energy-losses}).

\begin{figure}[!t]
  \centering
  \includegraphics{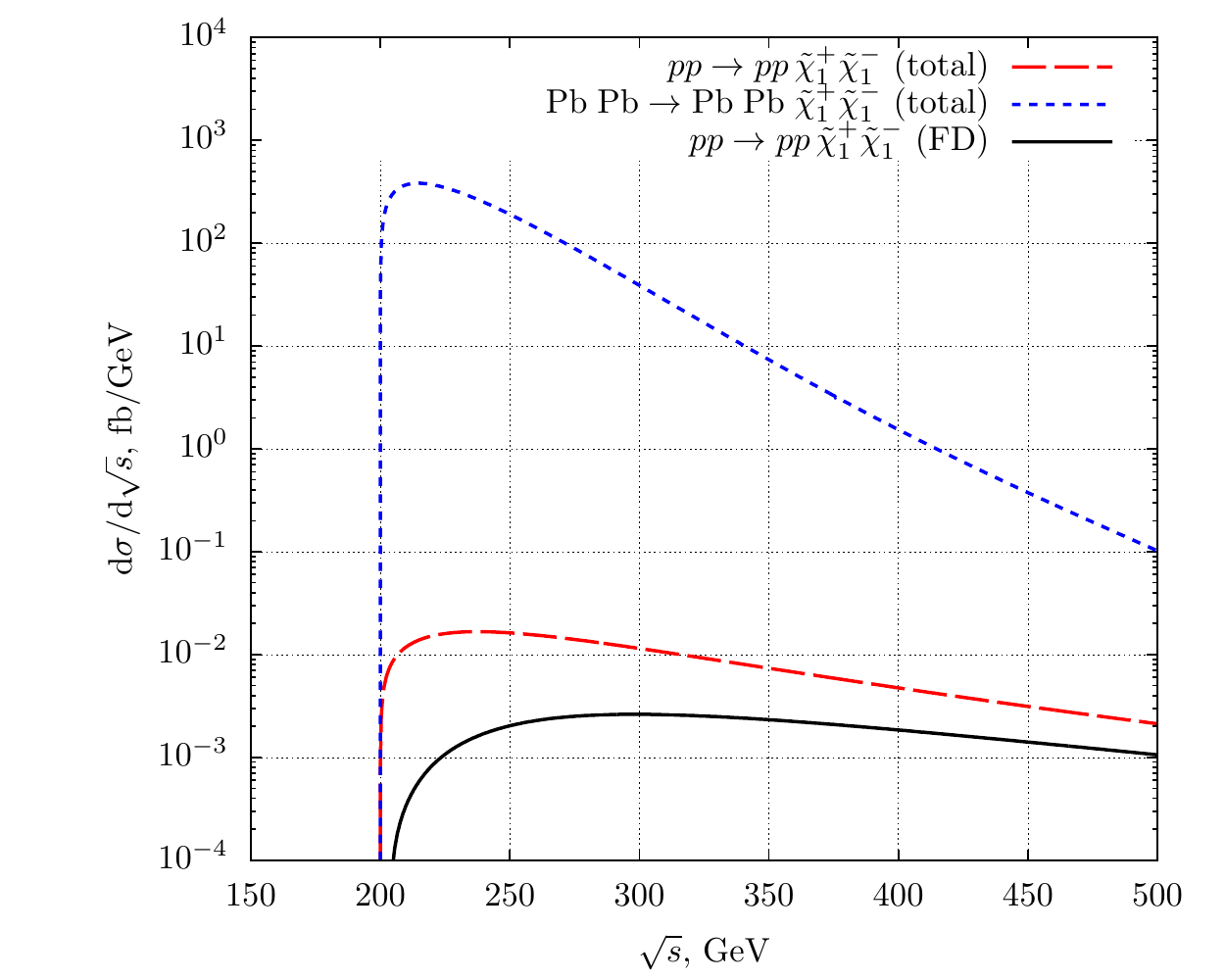}
  \caption{Differential cross sections for chargino pair production in
  ultraperipheral collisions at the LHC with respect to the chargino pair
  invariant mass $\sqrt{s}$ (often denoted as $\sqrt{s_{\gamma \gamma}}$ in
  other papers). $pp \to pp \ch^+ \ch^-$ (total) and $\text{Pb}
  \; \text{Pb} \to \text{Pb} \; \text{Pb} \; \ch^+ \ch^-$ (total) are the cross
  sections integrated over whole phase space. $pp \to pp \ch^+ \ch^-$ (FD) is
  the cross section with the requirement that both protons are detected in
  forward detectors (FD). The FD cross section for lead-lead collisions is
  many orders of magnitude less, and its threshold is at $\sqrt{s} \approx
  15$~TeV, because both lead ions have to lose at least $7.8$~TeV of their energy to hit
  forward detectors (see Table~\ref{t:energy-losses}). Here chargino mass is
  assumed to be 100~GeV, $pp$ collision energy is 13~TeV, Pb~Pb collision energy
  is $5.02~\text{TeV}/\text{nucleon pair}$.}
  \label{f:xsections}
\end{figure}

Differential cross sections are presented in
Fig.~\ref{f:xsections}. Assuming total Run~3 luminosity in proton-proton
collisions of 300~fb$^{-1}$ at the ATLAS and CMS detectors, the number of
produced chargino pairs with both protons detected in forward detectors can be
of the order of 250 per detector.

\begin{figure}[p]
  \centering
  \includegraphics{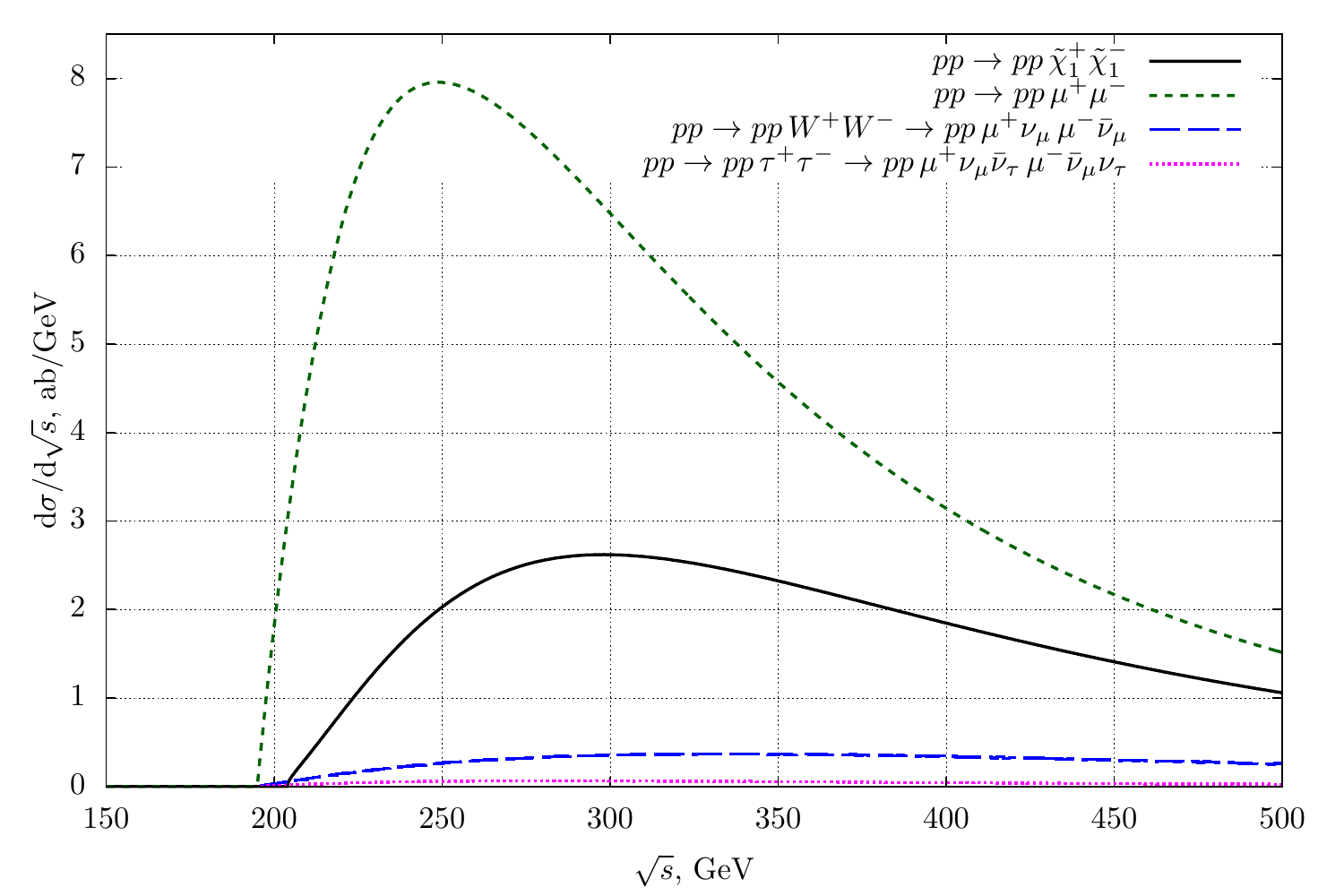}
  \caption{
     Differential fiducial cross sections for the reaction $pp \to pp \, \ch^+
     \ch^-$ and its backgrounds with respect to the invariant mass of the
     produced system.}
  \label{f:background}
\end{figure}

\begin{figure}[p]
  \centering
  \includegraphics{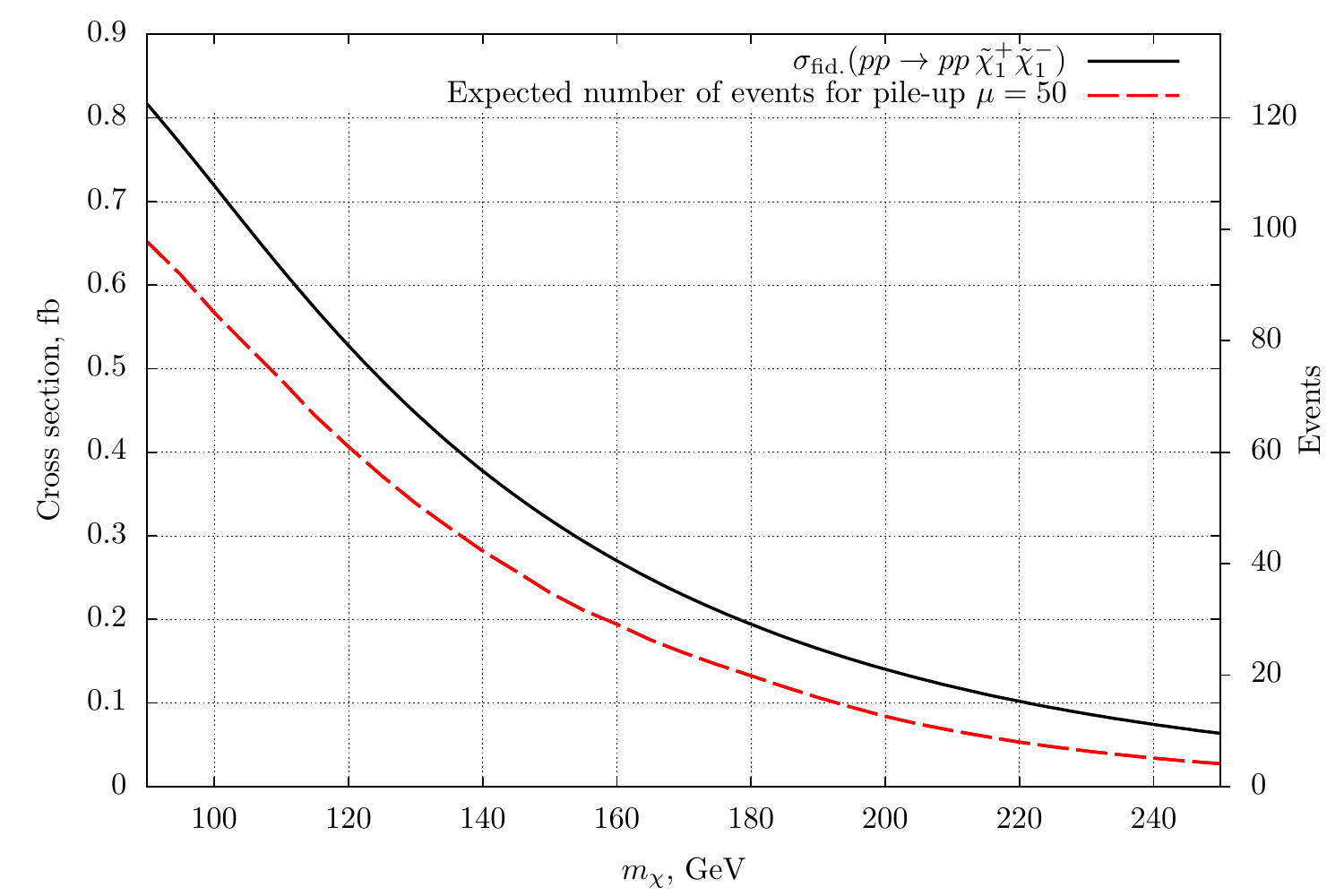}
  \caption{
    {\it Upper curve:} fiducial cross section for the $pp \to pp \; \ch^+ \ch^-$
    reaction with respect to chargino mass.
    {\it Lower curve:} expected number of events in the LHC data collected in
    Run~2 (for the integrated luminosity 150~fb$^{-1}$) assuming constant
    pile-up $\mu = 50$ and after applying the cut~\eqref{pl-cut}.
  }
  \label{f:xsection-m-fid}
\end{figure}

In the LHC experiments, some regions of the phase space of produced particles
are cut off.  Common requirements for a particle to be detected in the muon
system are $p_T > \hat p_T$ and $\abs{\eta} < \hat \eta$, where $p_T$ is the
particle transverse momentum, $\eta$ is its pseudorapidity, and $\hat p_T$ and
$\hat \eta$ are experimental cuts on these values. The corresponding (fiducial)
cross section for the $pp \to pp \, \ch^+ \ch^-$ reaction is (see
\cite{1806.07238} for the derivation of this formula with $m_\chi = 0$)
\begin{multline}
  \sigma_\text{fid.}(pp \to pp \, \ch^+ \ch^-) = \\
  = \int\limits_{(2 \xi_\text{min} E)^2}^{(2 \xi_\text{max} E)^2} \mathrm{d} s
    \int\limits_{
      \max \left(
        \hat p_T, \frac{\sqrt{s / 4 - m_\chi^2}}{\cosh \hat \eta}
      \right)
    }^{\sqrt{s / 4 - m_\chi^2}}
    \mathrm{d} p_T
    \, \frac{\mathrm{d} \sigma(\gamma \gamma \to \ch^+ \ch^-)}{\mathrm{d} p_T}
    \int\limits_{1/\hat x}^{\hat x} \frac{\mathrm{d} x}{8 x}
    \, n \left( \sqrt{\frac{sx}{4}} \right)
    \, n \left( \sqrt{\frac{s}{4x}} \right),
  \label{chargino-xsection}
\end{multline}
where $x = \omega_1 / \omega_2$, and
\begin{equation}
  \begin{aligned}
    \hat x
    &= \left( \hat X + \sqrt{\hat X^2 + 1} \right)^2, \\
    \hat X
    &= \frac{\sqrt{s} \, p_T}{2(p_T^2 + m_\chi^2)}
       \left(
         \sinh \hat \eta
         - \sqrt{\cosh^2 \hat \eta + \frac{m_\chi^2}{p_T^2}}
           \cdot
           \sqrt{1 - \frac{4 (p_T^2 + m_\chi^2)}{s}}
       \right).
  \end{aligned}
\end{equation}
The differential with respect to $p_T$ cross section is
\begin{equation}
  \frac{\mathrm{d} \sigma(\gamma \gamma \to \ch^+ \ch^-)}{\mathrm{d} p_T}
  = \frac{8 \pi \alpha^2 p_T}{s (p_T^2 + m_\chi^2)}
    \cdot
    \frac{1 - \dfrac{2 (p_T^4 + m_\chi^4)}{s (p_T^2 + m_\chi^2)}}
         {\sqrt{1 - \dfrac{4 (p_T^2 + m_\chi^2)}{s}}}.
\end{equation}
For
\begin{equation}
  \begin{aligned}
    m_\chi &= 100~\text{GeV}, & E &= 6.5~\text{TeV}, \\
    \xi_\text{min} E &= 97.5~\text{GeV}, & \xi_\text{max} E &= 975~\text{GeV}, \\
    \hat p_T &= 20~\text{GeV}, & \hat \eta &= 2.5,
  \end{aligned}
  \label{parameters}
\end{equation}
we get
\begin{equation}
  \sigma_\text{fid.}(pp \to pp \, \ch^+ \ch^-) = 0.72~\text{fb}.
\end{equation}
The differential fiducial cross section is presented in Fig.~\ref{f:background}.
Integrated fiducial cross section for a range of chargino masses is presented in
Fig.~\ref{f:xsection-m-fid}.

\section{Search strategy}

\label{s:search}

Assuming $R$-parity conservation, with the lightest chargino and the lightest
neutralino masses being nearly equal, it is possible that $\ch^\pm$ lives long
enough to escape the detector and decay outside.  The experimental signature
and, consequently, the background of the chargino production process greatly
depend on the stability of chargino. There are three possible scenarios:
\begin{minipage}{\textwidth} 
\begin{enumerate}
  \item Chargino decays in the beam pipe of the detector. This scenario will not
  be studied in this paper. For model-dependent bounds see,
  e.g.,~\cite{ATLAS-CONF-2019-014, 1905.13059}.
  \item Chargino decays in the body of the detector producing a disappearing track
  in the detector. This scenario is studied in Ref.~\cite{1712.02118,
  1804.07321} in the framework of minimal anomaly mediated symmetry breaking
  model (mAMSB), and charginos with the mass 100~GeV and lifetime above 0.02~ns
  and below few 100~ns are excluded. These searches require high energetic jet
  from initial state radiation in chargino production events, so their bounds
  are not applicable for chargino production in ultraperipheral collisions.
  This scenario will not be further studied in this paper.
  \item \label{i:stable-chargino} Chargino decays outside the detector producing
  a track in the detector.
\end{enumerate}
\end{minipage}

Let us consider the case when chargino lives long enough to escape the detector
(case~\ref{i:stable-chargino}). Then a track from a charged particle will be
observed in the detector. Since in the Standard Model only a muon can go through
the full detector (including the outer muon spectrometers), the question is
whether a chargino can be distinguished from a muon. The common approach for the
search for long-lived charged heavy particles is to measure their energy loss
($\mathrm{d} E / \mathrm{d} x)$ and time of flight through the detector
(TOF)~\cite{1101.1645, 1106.4495, 1205.0272, 1305.0491, 1411.6795, 1502.02522,
1506.05332, 1506.09173, 1604.04520, 1609.08382, 1808.04095, 1902.01636}. An
advantage of UPC is that the event kinematics can be fully reconstructed by
measuring the proton energies in forward detectors. In what follows we will
study the possibilities provided by this feature of UPC. The method proposed can
be complemented by the conventional $\mathrm{d} E / \mathrm{d} x$ and TOF
measurements.

For the reaction $pp \to pp \ch^+ \ch^-$, momenta of all four particles in the
final state can be measured: momenta of chargino candidates $\vec p_1$, $\vec
p_2$ can be reconstructed from their tracks in the detector, and final state
protons can be detected by the forward detectors thus providing their energy
losses $\xi_1$, $\xi_2$ (protons transverse momenta can be neglected). The
observable suitable for the discovery of chargino in UPC is the mass of the
charged particle
\begin{gather}
  m
   = \sqrt{
      \frac14
      \left(
        E (\xi_1 + \xi_2)
        + \frac{\vec p_1^{\; 2} - \vec p_2^{\; 2}}{E (\xi_1 + \xi_2)}
      \right)^2
      - \vec p_1^{\; 2}
     }
   = \frac{
       \sqrt{
         \left(
           E^2 (\xi_1 + \xi_2)^2 - ( \vec p_1^{\; 2} + \vec p_2^{\; 2} )
         \right)^2
         - 4 \vec p_1^{\; 2} \vec p_2^{\; 2}
       }
     }{2 E (\xi_1 + \xi_2)},
  \label{chi-mass-1}
  \\
  m   
   = \sqrt{
       \frac{
          (2 \xi_1 \xi_2 E^2 + \vec p_1 \vec p_2)^2
          - \vec p_1^{\; 2} \vec p_2^{\; 2}
       }{
        4 \xi_1 \xi_2 E^2 + (\vec p_1 + \vec p_2)^2
       }
     }.
  \label{chi-mass-2}
\end{gather}
Eqs.~\eqref{chi-mass-1} and~\eqref{chi-mass-2} are equivalent due to momentum
conservation law, however experimental uncertainties give different
contributions to these formulas, so both of them are useful when dealing with
experimental data. In what follows we will use~\eqref{chi-mass-2}, which is
less affected by finite detector resolution. Calculating the mass according
to~\eqref{chi-mass-2} for every event with exactly two charged tracks and two
protons detected in forward detectors, and plotting the number of such events
with respect to $m$, one should get $\delta(m - m_\chi)$ smeared with the
detector resolution.

\section{Background}

\label{s:background}

\subsection{Muons}

A long-lived chargino produces a signal in the detector very similar to that of
a muon. Therefore, the sources of the background are the reactions producing a
pair of muons. We consider the following processes:
\begin{enumerate}
  \item $pp \to pp \, \mu^+ \mu^-$.
  \item $pp \to pp \, W^+ W^- \to pp \, \mu^+ \nu_\mu \, \mu^- \bar \nu_\mu$.
  \item $pp \to pp \, \tau^+ \tau^- \to pp \, \mu^+ \nu_\mu \bar \nu_\tau \,
  \mu^- \bar \nu_\mu \nu_\tau$.
\end{enumerate}

Eq.~\eqref{chargino-xsection} with $m_\chi$ replaced with $m_\mu$ also works for
the $pp \to pp \, \mu^+ \mu^-$ reaction.\footnote{The muon mass can be neglected
as long as $m_\mu^2 \ll \hat p_T^2$.} Fiducial cross sections for the $pp \to pp
\, W^+ W^- \to pp \, \mu^+ \nu_\mu \, \mu^- \bar \nu_\mu$ and $pp \to pp \,
\tau^+ \tau^- \to pp \, \mu^+ \nu_\mu \bar \nu_\tau \, \mu^- \bar \nu_\mu
\nu_\tau$ reactions were calculated with the help of the Monte Carlo method.
Parameters of the calculation are defined in Eq.~\eqref{parameters}.
Cross section for the $\gamma \gamma \to W^+ W^-$ process is~\cite{npb228-285}
\begin{multline}
     \sigma(\gamma \gamma \to W^+ W^-) =
       \\
     = \frac{8 \pi \alpha^2}{m_W^2}
       \left[
           \left( 1 + \frac{3 m_W^2}{4 s} + \frac{12 m_W^4}{s^2} \right)
           \sqrt{1 - \frac{4 m_W^2}{s}}
         - \frac{3 m_W^4}{4 s^2}
           \left( 1 - \frac{2 m_W^2}{s} \right)
           \ln \frac{1 + \sqrt{1 - 4 m_W^2 / s}}{1 - \sqrt{1 - 4 m_W^2 / s}}
       \right].
\end{multline}
The results are presented in Fig.~\ref{f:background} and
Table~\ref{t:fiducials}.
\begin{table}[!b]
  \centering
  \caption{Fiducial cross sections for the $pp \to pp \, \ch^+ \ch^-$ reaction
  and its backgrounds.}
  \begin{tabular}{lc}
    \hline
    \hline
    \multicolumn{1}{c}{Reaction}
    & Cross section, fb
    \\ \hline
    $pp \to pp \, \ch^+ \ch^- \quad (m_\chi = 100~\text{GeV})$
    & $0.72$
    \\
    $pp \to pp \, \mu^+ \mu^-$
    & $1.60$
    \\
    $pp \to pp \, W^+ W^- \to pp \, \mu^+ \nu_\mu \, \mu^- \bar \nu_\mu$
    & $0.15$
    \\
    $pp \to pp \, \tau^+ \tau^- \to pp \, \mu^+ \nu_\mu \bar \nu_\tau \, \mu^-
    \bar \nu_\mu \nu_\tau$
    & $0.02$ \\
    \hline
    \hline
  \end{tabular}
  \label{t:fiducials}
\end{table}

\begin{figure}[!b]
  \vspace*{-3mm}

  \centering
  \includegraphics{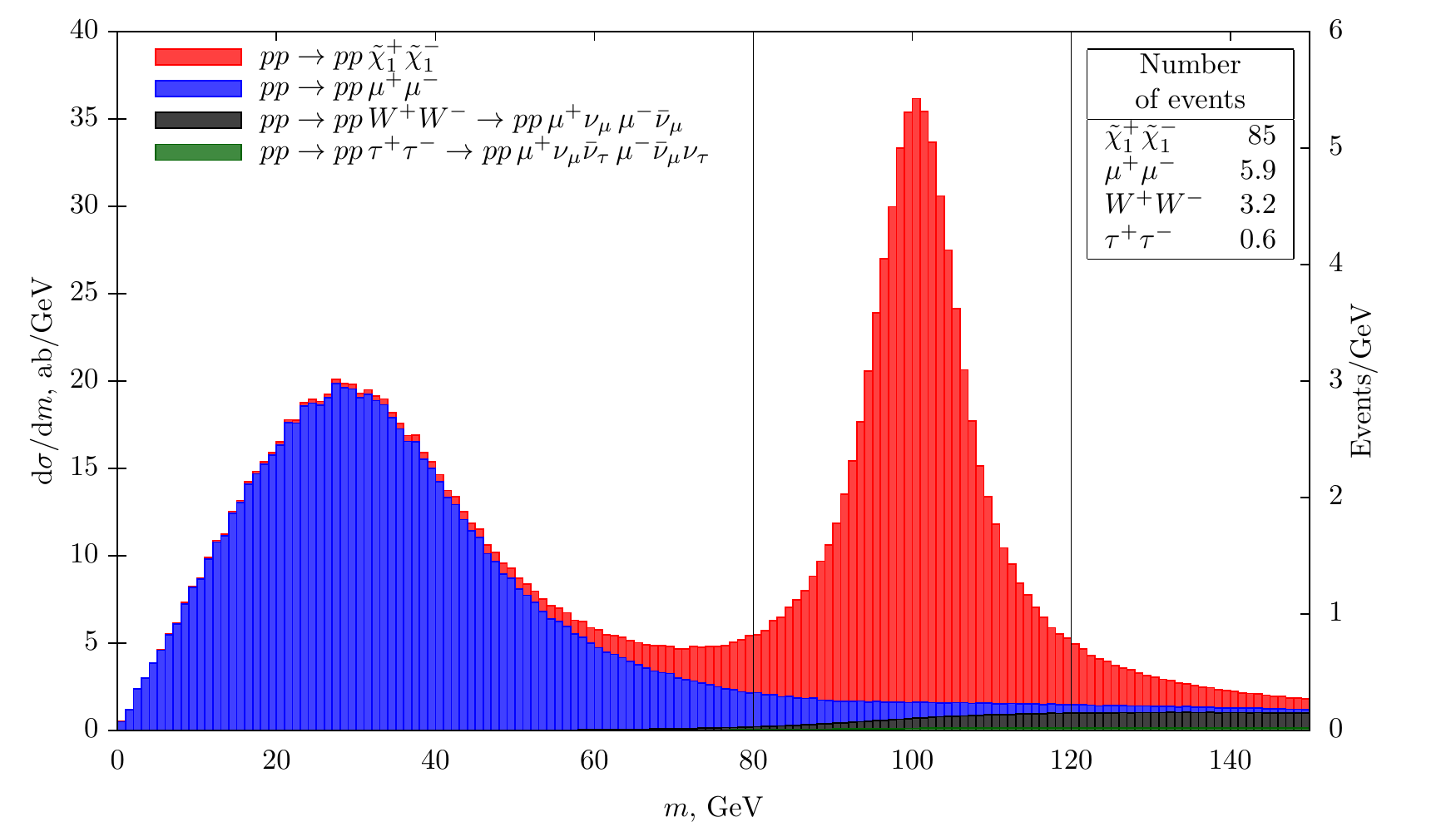}
  \caption{Monte Carlo simulation of the chargino candidate mass distribution.
  Here and in subsequent plots parameters of the calculation are defined in
  Eq.~\eqref{parameters}, the integrated luminosity is assumed to be
  150~fb$^{-1}$, bin width is 1 GeV and the values in the inset are the total
  number of the events in between the vertical lines.}
  \label{f:signal}
\end{figure}

Chargino candidate mass distributions according to Eq.~\eqref{chi-mass-2} for
the signal and background processes were calculated by means of the Monte Carlo
method. Finite central detector resolution was taken into account according
to~\cite[Section~4.5]{ATLAS-TDR-4}. Finite forward detector resolution was taken
into account by replacing in \eqref{chi-mass-2} $\xi_i E$ with a random number
normally distributed around $\xi_i E$ with the standard variation linearly
interpolated with pivot points 5~GeV for $\xi_i = 0.04$ and 10~GeV for $\xi_i =
0.14$, in accordance with~\cite[Section~3.3.2]{atlas-tdr-024-2015}.  The results
are presented in Fig.~\ref{f:signal}.  In the case of muons, in half of the
events $m^2$ is negative, and such events were discarded. When changing from the
distribution with respect to $m^2$ to the distribution with respect to $m$, an
extra factor of $m$ from the Jacobian results in the distribution being 0 at $m
= 0$ (see Appendix~\ref{a:jacobian}). This effect is mostly irrelevant for
charginos which peak is far from $m = 0$. The peak of charginos is  well
separated from the peak of muons. The background from $W^+ W^-$ and $\tau^+
\tau^-$ production and decay is negligible. Note that the background with
neutrinos in the final state can be further heavily suppressed by the
requirement that $p_{T1} + p_{T2} = 0$ where $p_{Ti}$ are transverse momenta of
the detected particles.

\subsection{Pile-up}

\label{s:pile-up}

Another important source of background is pile-up. During Run~2 of the LHC, the
pile-up was increasing from 25 to 38 collisions per bunch crossing on average,
reaching over 70 collisions in some events~\cite{atlas-luminosity,
cms-luminosity}. In what follows we consider a bunch crossing with $\mu =
50$ collisions. It is possible that in one of the collisions, a pair of muons is
produced with the energies high enough to pass the cuts on transverse momentum,
but not high enough for the proton(s) to be detected in the forward detector(s).
At the same time in one or more of the other 49 collisions, another event might
happen which results in the proton hitting the forward detector yet the produced
particles (e.g., pions) do not pass the trigger thresholds or escape the
detector through the beam pipe. In this case the muons from the original event
and the protons from the pile-up will mimic the production of a chargino pair
with relevant chargino masses.

The most probable event resulting in a proton hitting the forward detector is
proton diffractive dissociation which produces a few low-energy
pions which escape detection~\cite{prep50-157}.  Proton diffractive dissociation
is described by the triple-Regge diagrams~\cite{plb45-493}. In Appendix~B of
Ref.~\cite{1812.04886}, the probability for a proton after dissociation to hit
the forward detector was estimated to be $P_\text{SD} (1) \approx 0.01$ for
$0.02 < \xi < 0.15$.  The probability to observe one or more of such protons in
a single bunch crossing is
\begin{equation}
  P_\text{SD}(\mu) = 1 - (1 - P_\text{SD}(1))^\mu.
\end{equation}
$P_\text{SD}(50) \approx 0.39$, or, in other words, about 40\% of bunch
crossings with 50 collisions at once produce at least one proton hitting one of
the forward detectors. Following Ref.~\cite{1702.05023}, we shall use the
low-$\xi$ approximation for the differential cross section of proton
dissociation in the form
\begin{equation}
  M_X^2 \frac{\mathrm{d} \sigma}{\mathrm{d} M_X^2}
  \propto 1 + \frac{2~\text{GeV}}{M_X},
\end{equation}
where $M_X$ is the invariant mass of the system produced. Since $\xi = M_X^2 / 4
E^2$, this expression can be rewritten as
\begin{equation}
  \xi \frac{\mathrm{d} \sigma}{\mathrm{d} \xi}
  \propto 1 + \frac{\varkappa}{\sqrt{\xi}},
\end{equation}
where $\varkappa = 2~\text{GeV} / \, 13~\text{TeV} = 1.5 \cdot 10^{-4}$.
The corresponding spectrum of dissociated protons hitting the
forward detector per bunch crossing is
\begin{equation}
  f_p(\mu, \xi)
  = P_\text{SD}(\mu) \cdot
    \frac{
      \, \mathrm{d} \sigma / \mathrm{d} \xi
    }{
      \int\limits_{\xi_\text{min}}^{\xi_\text{max}}
        \frac{\mathrm{d} \sigma}{\mathrm{d} \xi}
        \, \mathrm{d} \xi
    }
  = \frac{
      P_\text{SD}(\mu) \cdot \dfrac{1}{\xi}
      \left( 1 + \dfrac{\varkappa}{\sqrt{\xi}} \right)
    }{
      \ln \dfrac{\xi_\text{max}}{\xi_\text{min}}
      - \dfrac{\varkappa}{2}
        \left(
          \dfrac{1}{\sqrt{\xi_\text{max}}} - \dfrac{1}{\sqrt{\xi_\text{min}}}
        \right)
    }.
\end{equation}

\begin{figure}[!tbhp]
  \centering
  \includegraphics{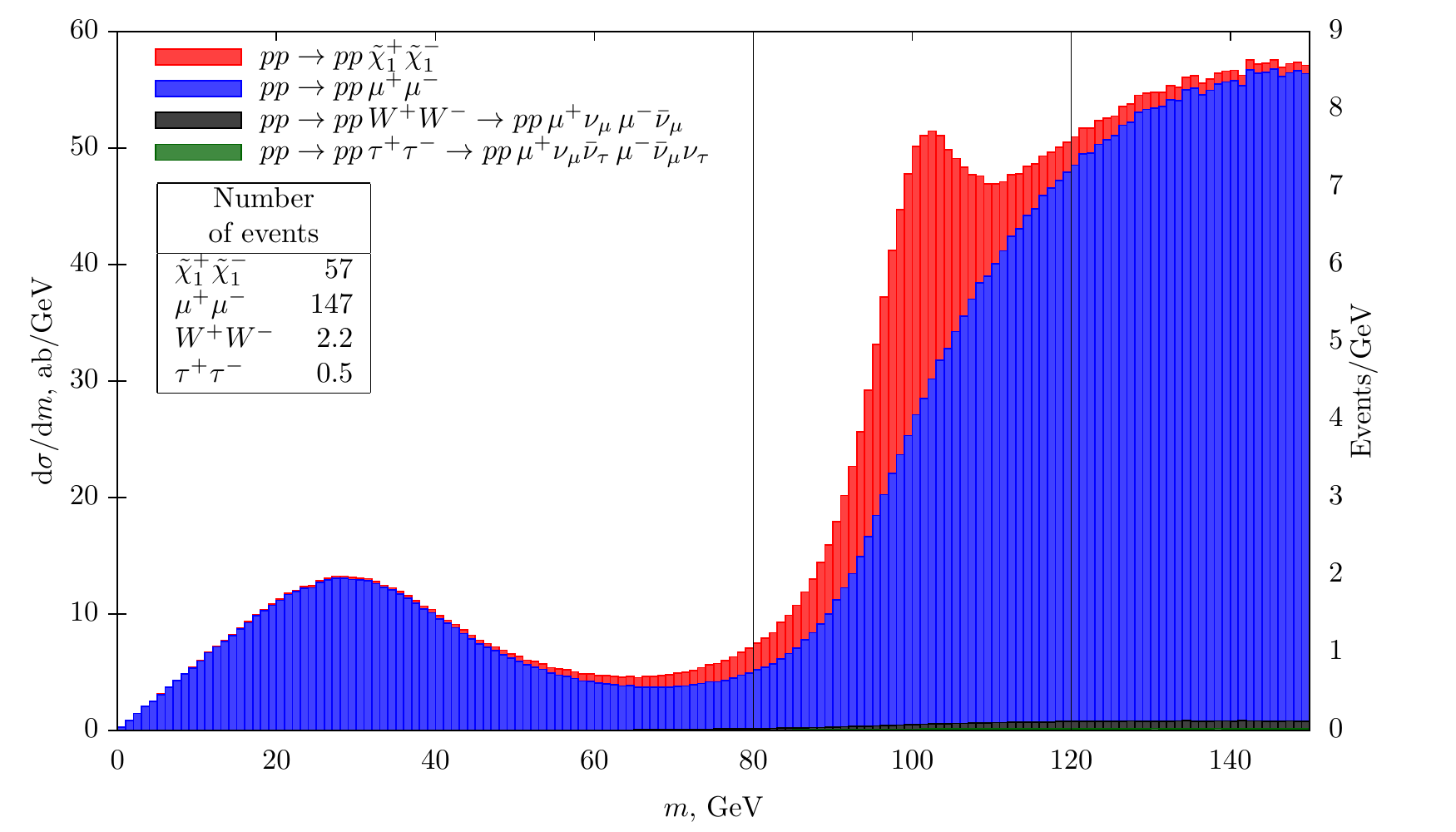}
  \caption{Monte Carlo simulation of the chargino candidate mass distribution
  with the pile-up background ($\mu = 50$).}
  \label{f:pu-nocut}
\end{figure}

The results of Monte Carlo simulation with pile-up value $\mu = 50$ are
presented in Fig.~\ref{f:pu-nocut}. In this case the background is much larger
than the signal. The reason is that the equivalent photon spectrum increases at
low photon energies, so production of a pair of muons in an UPC is much more
probable than production of a pair of charginos with the mass 100~GeV. With no
pile-up events, the background from muons was suppressed by the lower bound of
the forward detector acceptance region $\xi_\text{min}$: if the invariant mass
of the muon pair was less than $2 \xi_\text{min} E = 195$~GeV, both protons
could not hit the forward detectors simultaneously. With the pile-up enabled,
one or two of the protons can come from the pile-up.  In Fig.~\ref{f:pu-nocut}
events with multiple hits in forward detectors were discarded.

The advantage of ultraperipheral collisions in the case of pair production of
quasistable charginos is that all particles in the final state can be detected
and their momenta can be measured. This information can be used to greatly
suppress the pile-up background. The total 3-momentum of the colliding system is
zero. Its longitudinal component after the collision,
\begin{equation}
  p_{\parallel, 1} + p_{\parallel, 2} + (1 - \xi_1) E - (1 - \xi_2) E = 0,
\end{equation}
where $p_{\parallel, 1}$, $p_{\parallel, 2}$, $(1 - \xi_1) E$, and $-(1 -
\xi_2)E$ are longitudinal components of momenta of the charginos and the
protons.  In the case of the pile-up, one or both of the protons are produced in
a different event, and this equation is violated. Hence, the background is
suppressed by the cut
\begin{equation}
  \abs{p_{\parallel, 1} + p_{\parallel, 2} - (\xi_1 - \xi_2) E}
  < \hat p_\parallel.
  \label{pl-cut}
\end{equation}
This cut also suppresses the background from $W^+ W^-$ and $\tau^+ \tau^-$
production since neutrinos carry away longitudinal momentum.

\begin{figure}[!tbhp]
  \centering
  \includegraphics{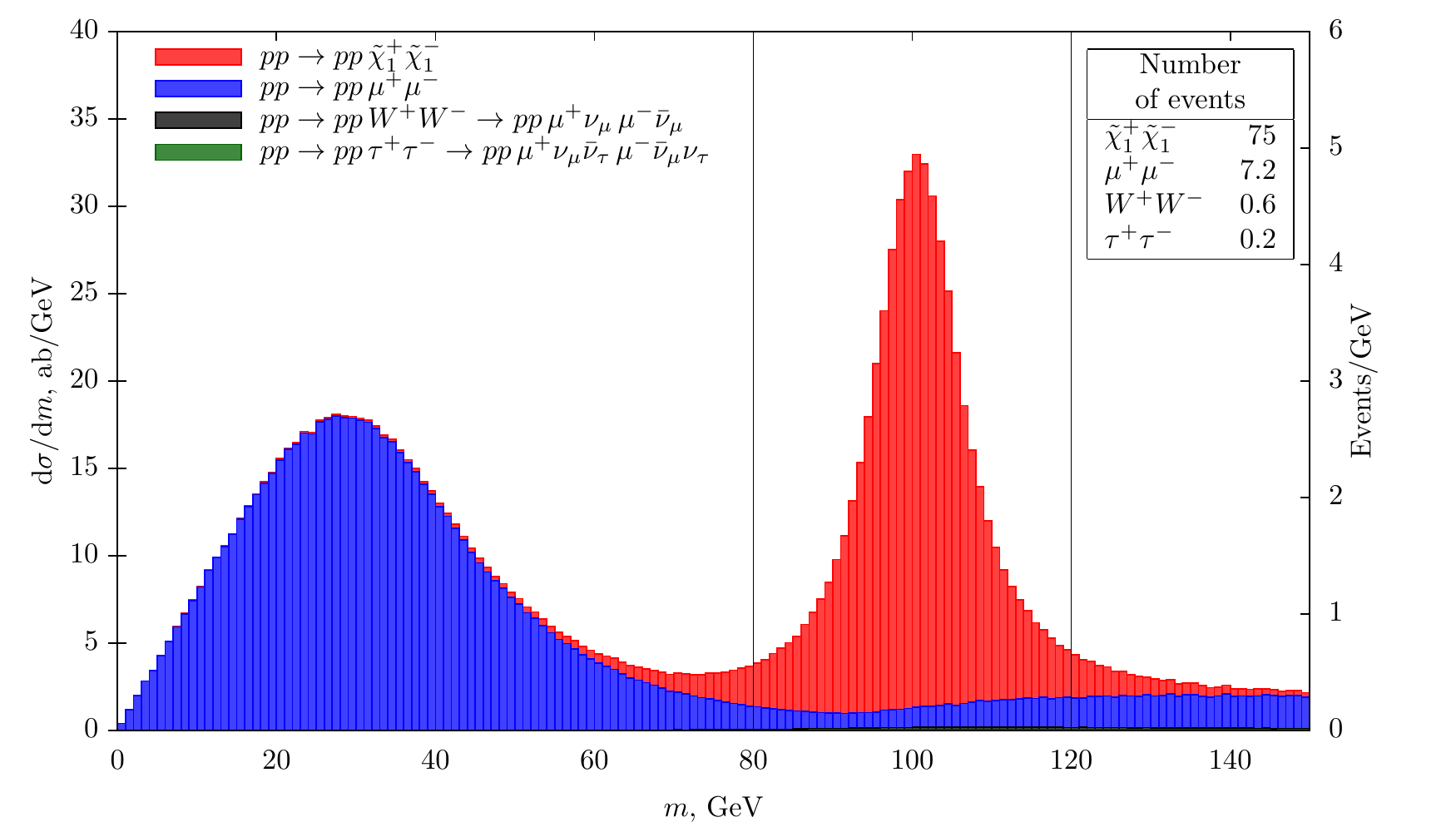}
  \caption{Monte Carlo simulation of the chargino candidate mass distribution
  with the pile-up background and the cut on the longitudinal momentum of the
  final state system~\eqref{pl-cut}.}
  \label{f:pileup}
\end{figure}

Chargino mass distribution for the same parameters as in Fig.~\ref{f:signal}
with the pile-up value $\mu = 50$ was calculated with the help of the Monte
Carlo method. The value of $\hat p_\parallel$ was chosen to be 20~GeV. The
result is presented in Fig.~\ref{f:pileup}. In the case of multiple hits in
forward detectors, the pair of protons which satisfies Eq.~\eqref{pl-cut} was
selected (events with more than one such pair were discarded). This restores the
signal events that were discarded due to pile-up in Fig.~\ref{f:pu-nocut} almost
to the number that was available with no pile-up in Fig.~\ref{f:signal}. The
background from $W^+ W^-$ and $\tau^+ \tau^-$ production is less than
$0.15$~ab/GeV.

The total number of events expected to be observed in Run~2 data depending on
chargino mass is shown in Fig.~\ref{f:xsection-m-fid}.

\section{Accessible masses and lifetimes}

\label{s:lifetimes}

So far we were considering charginos living long enough to escape the detector.
However, chargino can decay in the detector reducing the number of events that
can be reconstructed with our method. In this section we will estimate the range
of chargino lifetimes that would allow for the observation of charginos with the
LHC data collected during Run~2 assuming that forward detectors operated at
100\% efficiency.

\begin{figure}[!tbhp]
  \centering
  \includegraphics{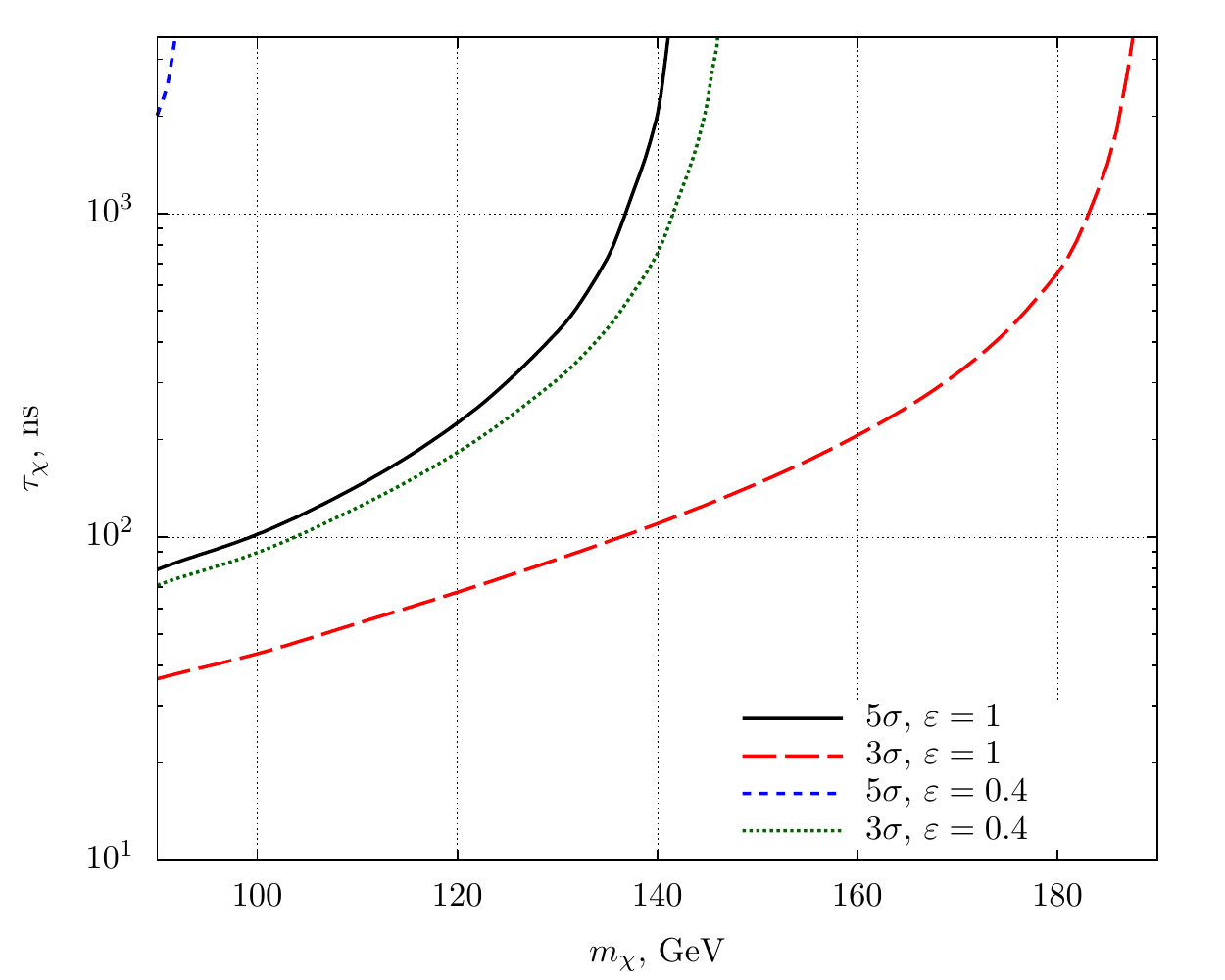}
  \caption{Lower bounds on the values $(m_\chi, \tau_\chi)$ that could be
  observed with the LHC Run~2 data. $3 \sigma$ and $5 \sigma$ are the signal
  significances. $\varepsilon$ is the trigger and reconstruction efficiencies.
  Here ATLAS detector geometry and resolution were used.}
  \label{f:lifetimes}
\end{figure}

\begin{figure}[!tbhp]
  \centering
  \includegraphics{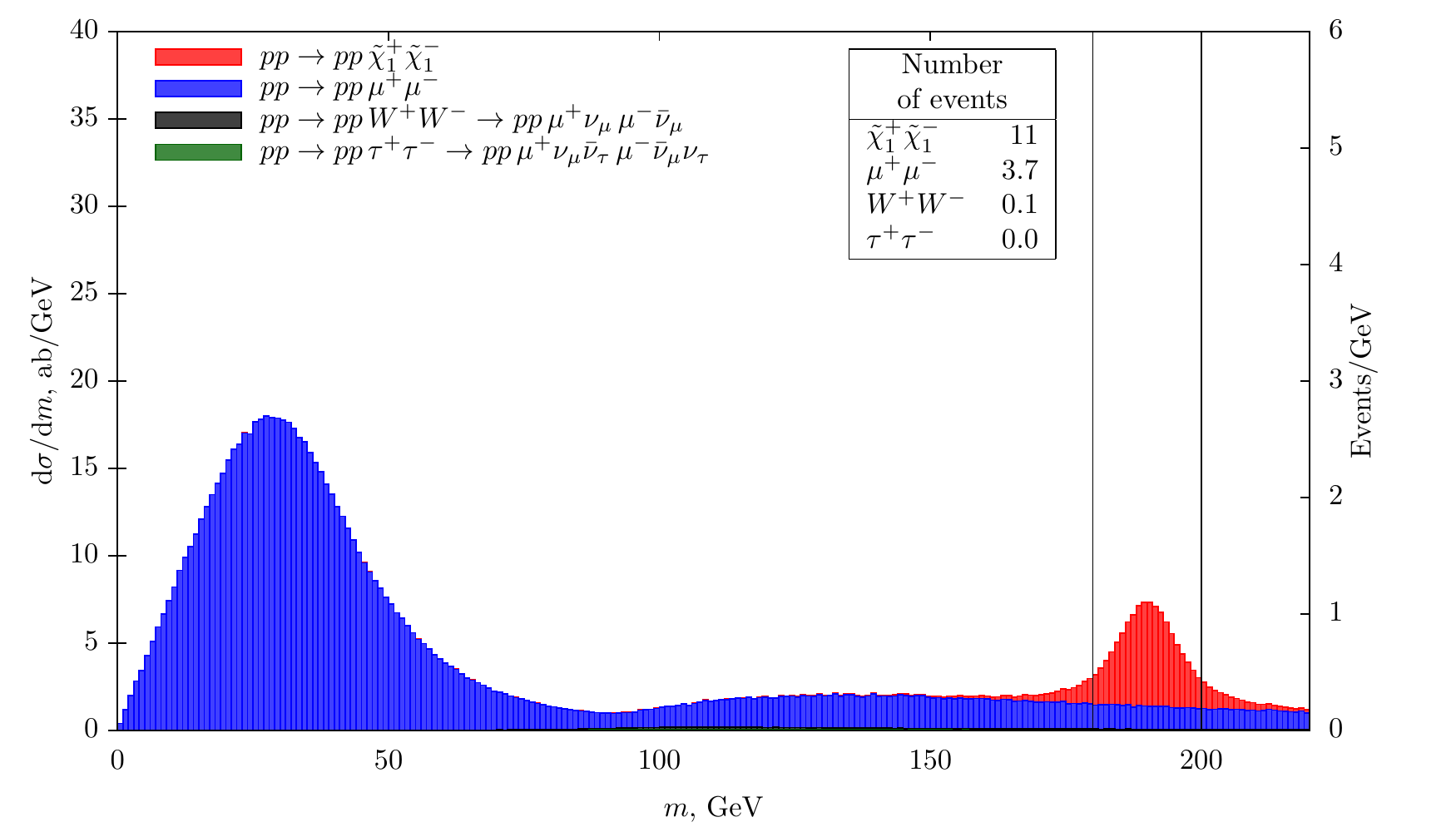}
  \caption{Same as Fig.~\ref{f:pileup}, but with chargino mass set to 190~GeV.}
  \label{f:pileup-2}
\end{figure}

For that purpose we have implemented the possibility for chargino decay in our
Monte Carlo simulations. Both charginos have to escape the detector for the
event to be selected. Chargino candidate mass distributions were calculated for
a set of chargino masses $m_\chi$ and lifetimes $\tau_\chi$ for the pile-up $\mu
= 50$. For each distribution the mass range $m_\chi - 10~\text{GeV} < m < m_\chi
+ 10~\text{GeV}$ was selected and the corresponding number of signal events
$n_\chi$ from the $pp \to pp \ch^+ \ch^-$ reaction and the total number of
events $n$ including the background processes were calculated. Significance was
estimated as $S = n_\chi / \sqrt{n}$. Fig.~\ref{f:lifetimes} shows isolines of
$S$ corresponding to $S = 3$ and $S = 5$. The largest chargino mass that can be
observed at the level of 3 standard deviations is 190~GeV.  Chargino candidate
mass distribution for stable chargino with the mass 190~GeV is presented in
Fig.~\ref{f:pileup-2}. In the region of masses from 180 to 200 GeV the signal is
11 events, the background is $3.7$ events.

Our results do not take into account the reduction of the number of events due
to trigger and reconstruction efficiency. In Ref.~\cite{1708.04053}, the
reconstruction efficiency for the $pp \to pp \mu^+ \mu^-$ reaction was estimated
at the level of $\varepsilon = 0.4$. To demonstrate the effect of efficiency, we
have reduced the number of events correspondingly and presented the isolines of
$S$ for this value in Fig.~\ref{f:lifetimes} as well.

Decaying charginos will leave disappearing tracks in the detector. If the
chargino lifetime is too small for chargino to be observed with the method
proposed, the latter can be complemented by the search for disappearing tracks.
Existing searches for disappearing tracks~\cite{1712.02118, 1804.07321} require
a jet in the final state which will not be present in the case of
ultraperipheral collisions.

\section{Conclusions}

Ultraperipheral collisions of protons allow production of heavy charged
particles in very clean events. Its cross section is unambiguously determined
by the electric charge and the mass of the particles produced and it does not
depend on the particular model of New Physics. Measurement of momenta of heavy
long-lived charged particles in the main detector complemented by the
measurement of protons momenta in forward detectors allows complete
reconstruction of event kinematics. With the LHC Run~2 data the range of
particle masses available for observation at the level of 3 standard deviations
reaches 190~GeV (see Figs.~\ref{f:xsection-m-fid} and~\ref{f:pileup-2}).
Accessible masses and lifetimes are shown in Fig.~\ref{f:lifetimes}. Particles
with the lifetime greater than 100~ns can be observed with the method proposed.
In view of these results, the operation of the ATLAS and CMS forward detectors
during Run 3 will allow for the discovery of new heavy quasistable charged
particles provided they exist.

Ultraperipheral lead-lead collisions provide orders of magnitudes greater
production cross section than the proton-proton collisions (see
Fig.~\ref{f:xsection-m}). However, complete reconstruction of kinematics with
the help of forward detectors is impossible in the case of lead-lead collisions
because the required energy loss is too large. New particles can still be
searched for with conventional methods based on the ionization energy loss and
time-of-flight measurements. These measurements will benefit from the very clean
final state of ultraperipheral collisions even in the case of collisions of lead
ions. In order to look for long-lived charged particles in lead-lead collisions,
the integrated luminosity should be increased by 2--3 orders of magnitude (see
the text after Eq.~\eqref{epa-xsection:fd}). This call for significantly larger
future heavy ion runs and\slash{}or runs with lighter ions is discussed in
detail in Ref.~\cite{1812.07688}.

\section{Acknowledgments}

We are grateful to K.~G.~Boreskov, A.~D.~Stepennov and I.~I.~Tsukerman for
useful discussions, and to V.~A.~Khoze for bringing to our attention the large
pile-up background. We thank the unknown referee for the comments that helped
to substantially improve the paper. The authors are supported by the Russian
Science Foundation grant No 19-12-00123.

\appendix

\section{On preceding searches for heavy charged long-lived particles}

\label{s:llp-searches}

If heavy long-lived charged particles (LLCP) exist and have masses
100--200~GeV, they would have already been produced in ultraperipheral
collisions at the LHC. They could be observed in searches which do not require
large missing energy or jets in the final state. Most of Refs.~\cite{1101.1645,
1106.4495, 1205.0272, 1305.0491, 1411.6795, 1502.02522, 1506.05332, 1506.09173,
1604.04520, 1609.08382, 1808.04095, 1902.01636} look for events with two
well-separated muon-like tracks with large transverse momentum not accompanied
by hadronic jets. We shall discuss why these studies do not exclude LLCP with
the masses 100--200 GeV produced in $pp$ UPC.

Ref.~\cite{1101.1645} is devoted to the search for heavy stable charged
particles in $pp$ collisions at the collision energy 7~TeV. Production cross
section in UPC~\eqref{epa-xsection} at this energy is $1.19$~fb for chargino
mass 100~GeV, and it decreases when the mass increases.  With the integrated
luminosity studied in this paper ($3.1~\text{pb}^{-1}$), no events are expected.

In Ref.~\cite{1106.4495}, the collision energy is 7~TeV as well, but the
integrated luminosity is about one order of magnitude larger
($37~\text{pb}^{-1}$). However, the luminosity is still too low for even a
single  $\ch^+ \ch^-$ pair to be produced.

In Ref.~\cite{1205.0272}, $5~\text{fb}^{-1}$ of integrated luminosity in 7~TeV
$pp$ collisions are analyzed. According to~\eqref{epa-xsection}, 6 events of
chargino pair production are expected in this data. For an event to be selected,
transverse momentum of each LLCP was required to be larger than $\hat p_T =
40$~GeV and its pseudorapidity to be less than $\hat \eta = 2.1$. With the help
of Eq.~\eqref{chargino-xsection}, setting $\xi_\text{min} = 0$, $\xi_\text{max}
= 1$ (so each proton can miss the forward detector), we get the corresponding
fiducial cross section of $0.65$~fb. 3 events passing these experimental cuts
are expected, and even no observation is compatible with this expectation.
Additional cuts implemented in Ref.~\cite{1205.0272} will further diminish the
number of events.

In Ref.~\cite{1305.0491}, $5.0~\text{fb}^{-1}$ of data at the collision energy
7~TeV and $18.8~\text{fb}^{-1}$ of data at the collision energy 8~TeV are
analyzed. The selection criteria for LLCP are $p_T > 70$~GeV, $\abs{\eta} <
2.1$. The strongest bound comes from the analysis taking both the $\mathrm{d}
E/\mathrm{d} x$ and time-of-flight (TOF) measurements into account. As is shown
in Table~3 of that paper, the expected number of background events for $m_\chi >
100$~GeV is $1.0 \pm 0.2$ for the collision energy 7~TeV and $5.6 \pm 1.1$ for
the collision energy 8~TeV.  The number of observed events are 3 and 7
correspondingly. Cross section for pair production of LLCP with the mass 100~GeV
in UPC passing the selection criteria is $0.47$~fb ($0.57$~fb) for 7 (8)~TeV
which corresponds to $2.3$ ($10.7$) events. To compare these numbers with the
number of observed events, the former should be multiplied by the detector
efficiency. From Table~3 and Figure~8 of the paper we estimate the detector
efficiency to be less than 20\% for $m_\chi = 100$~GeV. The resulting number of
events is compatible with the background fluctuation.

In Ref.~\cite{1411.6795}, $19.1~\text{fb}^{-1}$ of integrated luminosity in
8~TeV $pp$ collisions are analyzed. The relevant signal region is
\texttt{SR-CH-2C} which considers chargino pair production with two charged
tracks observed in the detector. The cuts on the phase space are $p_T > 70$~GeV,
$\abs{\eta} < 2.5$. Chargino mass is varied in the interval 100--800~GeV, but
the results are presented only for $m_\chi > 450$~GeV. As is shown in Table~6,
for $m_\chi = 500$~GeV, the detector efficiency is $0.061 \pm 0.003$.  The cross
section for production of a pair of LLCPs with the mass 100~GeV in UPC is
$0.66$~fb, so the number of expected events in this amount of data is $12.6$.
Assuming that the detector efficiency does not exceed 10\% for $m_\chi =
100$~GeV, we obtain that only about 1 event is expected. No events observed is
consistent with this expectation.

Ref.~\cite{1502.02522} sets bounds on pMSSM and AMSB models using the results of
Ref.~\cite{1305.0491}.

Ref.~\cite{1506.05332} considers production of metastable charged particles
which decay inside the detector. Only the events with the missing energy
$E_T^\text{miss} > 80$~GeV were selected. There is no missing energy in pair
production of LLCP in UPC, so such events would not be seen in this analysis.

In Ref.~\cite{1506.09173}, the LHCb collaboration searches for the LLCP with the
ring imagining Cherenkov detectors. 1~fb$^{-1}$ of data collected at $pp$
collisions with the energy 7~TeV and 2~fb$^{-1}$ of data collected at $pp$
collisions with the energy 8~TeV are used in this analysis. For an event to be
selected, both LLCP candidates have to have $p_T > 50$~GeV and hit the detector
located at $1.8 < \eta < 4.9$. Cross section for pair production of LLCP with
the mass $m_\chi = 100$~GeV in UPC of protons with the energy 8~TeV satisfying
the selection criteria is definitely less than $0.45$~fb. The predicted number
of events is less than 1. The background from the $Z/\gamma^* \to \mu^+ \mu^-$
reaction is much larger than 1.  Therefore, production of LLCP in UPC cannot be
observed in this experiment with the current amount of data.

In Ref.~\cite{1604.04520}, the ATLAS collaboration searches for long-lived
$R$-hadrons. As in Ref.~\cite{1506.05332}, large missing energy
($E_T^\text{miss} > 70$~GeV) is required, so this search is not sensitive to
ultraperipheral collisions.

In Ref.~\cite{1609.08382}, $2.5~\text{fb}^{-1}$ of $pp$ collisions with the
energy 13~TeV are analyzed. Cross section for pair production of LLCP with the
mass 100~GeV in UPC at this energy is $2.84$~fb~\eqref{epa-xsection:values}, so
$7.1$ events are expected in the data.  For the analysis, the events with $p_T >
55$~GeV and $\abs{\eta} < 2.1$ were selected. The UPC cross section diminishes
to $1.21$~fb which corresponds to $3.0$ events. As is shown in Table~1 of that
paper, 4 events were observed with the predicted number of background events
$5.4 \pm 1.1$. Even not taking into account the detector efficiency (30--40\%,
see Tables~5, 6), LLCP pair production in ultraperipheral collisions cannot be
excluded.

In Ref.~\cite{1808.04095}, as in Ref.~\cite{1604.04520}, long-lived $R$-hadrons
are searched for. Large missing energy (over 170~GeV) is required for an event
to be selected. This search is not sensitive to ultraperipheral collisions.

In Ref.~\cite{1902.01636}, $36.1~\text{fb}^{-1}$ of integrated luminosity in
proton-proton collisions with the energy of 13~TeV is studied. Several signal
regions are considered in the paper. The most sensitive region for LLCP pair
production in UPC is \texttt{SR-2Cand-FullDet} which requires two tracks from
charged particles with $p_T > 70$~GeV and $\abs{\eta} < 2.0$. The corresponding
fiducial cross section for $m_\chi = 100$~GeV is $0.97$~fb.  However, the region
$m_\chi < 200$~GeV was used as a control region in the paper. The fiducial cross
section in UPC for $m_\chi = 200$~GeV is $0.14$~fb which corresponds to $5.1$
events. The total cross section (not taking the experimental cuts into account)
is $0.27$~fb. According to Table~8, the product of detector acceptance and
efficiency is equal to $0.083 \pm 0.003$ for the process considered in the
paper. Assuming that the detector acceptance for LLCP pair production in UPC is
approximately the same as in the process considered in the paper, the detector
efficiency is $\sim 0.083 / (0.14 / 0.27) \approx 16$\%, so only $0.8$ events of
LLCP pair production in UPC are expected in this study.

We see that the cross sections of LLCP pair production in UPC convoluted with
actual detection efficiencies and luminosities are too small for these particles
to be detected in Refs.~\cite{1101.1645, 1106.4495, 1205.0272, 1305.0491,
1411.6795, 1502.02522, 1506.05332, 1506.09173, 1604.04520, 1609.08382,
1808.04095, 1902.01636}.

Due to pile-up, an ultraperipheral collision in which a LLCP pair is produced
may be accompanied by another collision with large missing energy. However,
taking into account that the number of charginos produced in UPC in the papers
considered above are of the order of 10, while cross section of the events with
large missing energy is many orders of magnitudes less than the total $pp$ cross
section ($\sim 100$~mb), the probability of such a coincidence is negligible.

\section{The shape of the reconstructed mass distribution}
\label{a:jacobian}

The reader may still be puzzled why the maximum of the reconstructed mass
distribution is at the value $\sim 30~\text{GeV}$ for muon background rather
than at the muon mass. This happens because we plot the reconstructed mass
distributions while the width of the squared mass distribution is much larger
than its mean value, i.e. the squared muon mass. The value of $\sim
30~\text{GeV}$ is defined by detector resolution and the energy spectrum of the
muon background.

\begin{figure}[p]
  \centering
  \begin{subfigure}[b]{0.493\textwidth}
    \centering
    \includegraphics[width=3.3in]{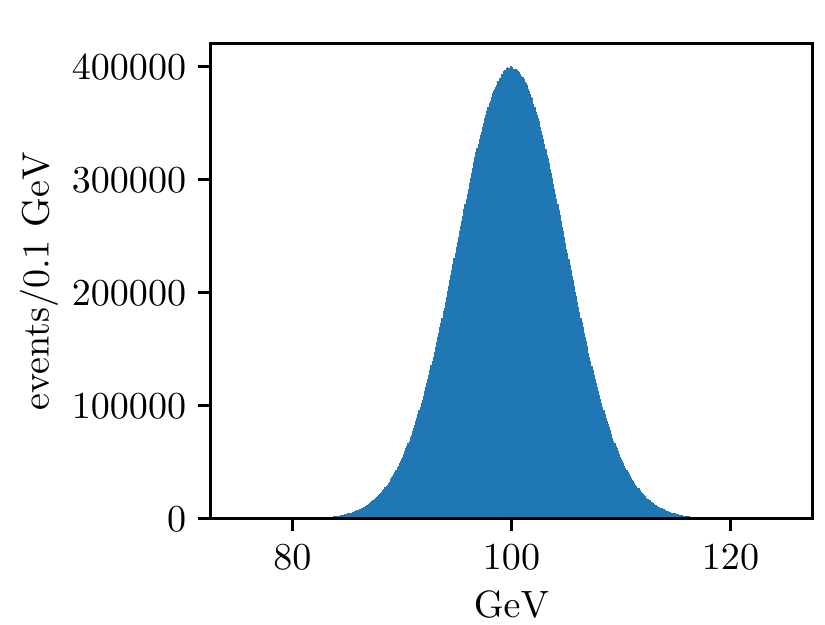}
    \caption{$E_{\rm rec}$.}
    \label{fig:E_rec}
  \end{subfigure}
  \begin{subfigure}[b]{0.493\textwidth}
    \centering
    \includegraphics[width=3.3in]{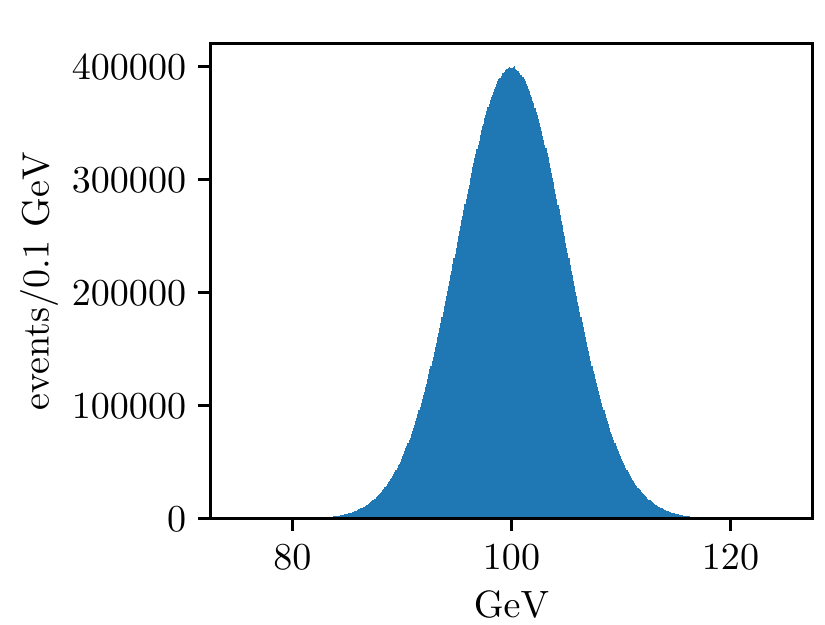}
    \caption{$p_{\rm rec}$.}
    \label{fig:p_rec}
  \end{subfigure}\\
  \begin{subfigure}[b]{0.493\textwidth}
    \centering
    \includegraphics[width=3.3in]{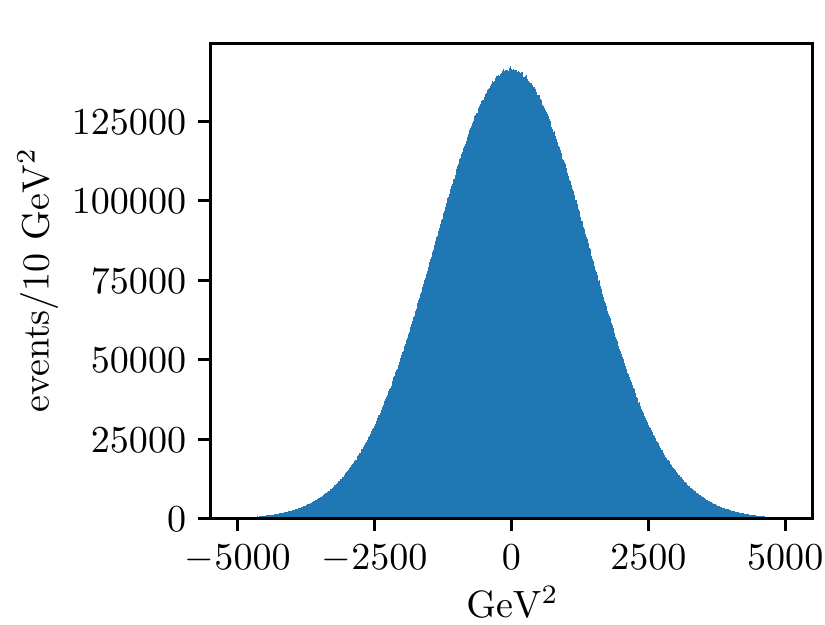}
    \caption{$m^{2}_{\rm rec}$.}
    \label{fig:m2_rec}
  \end{subfigure}
  \begin{subfigure}[b]{0.493\textwidth}
    \centering
    \includegraphics[width=3.3in]{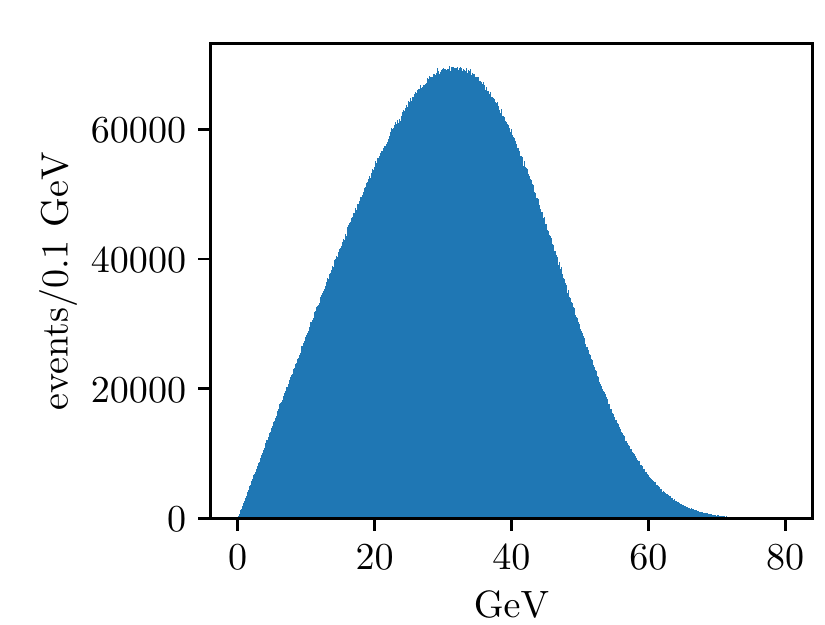}
    \caption{$m_{\rm rec}$.}
    \label{fig:m_rec}
  \end{subfigure}\\
  \begin{subfigure}[b]{0.493\textwidth}
    \centering
    \includegraphics[width=3.3in]{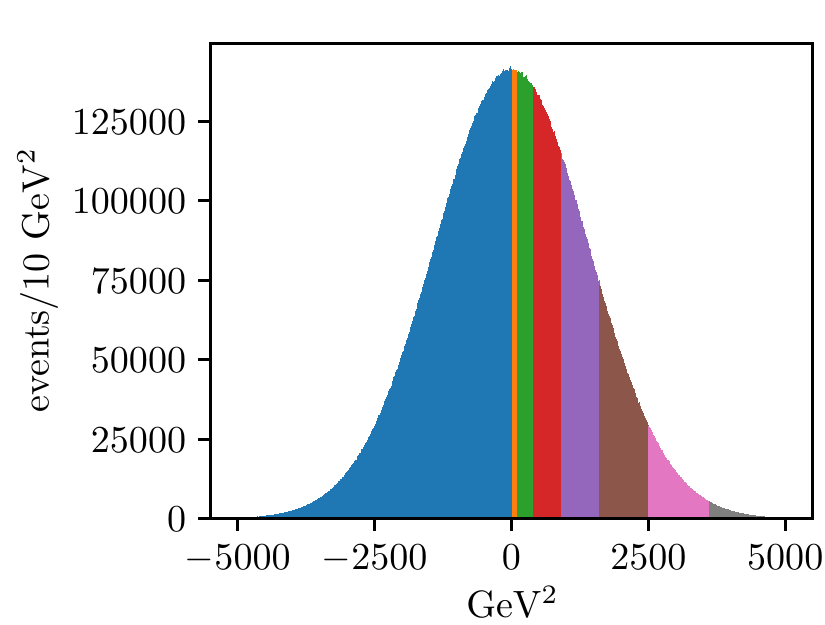}
    \caption{Highlighted $m^{2}_{\rm rec}$.}
    \label{fig:m2_rec_highlight}
  \end{subfigure}
  \begin{subfigure}[b]{0.493\textwidth}
    \centering
    \includegraphics[width=3.3in]{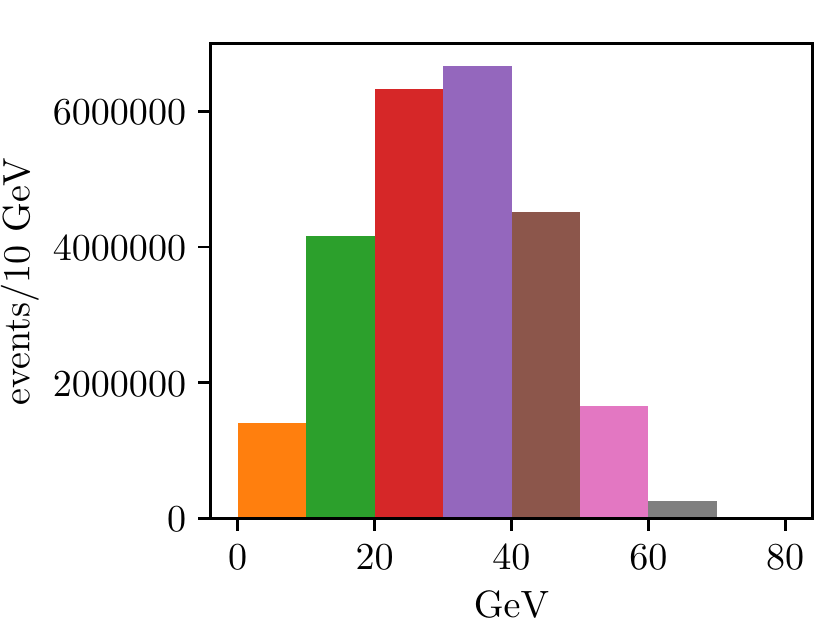}
    \caption{Highlighted $m_{\rm rec}$.}
    \label{fig:m_rec_highlight}
  \end{subfigure}
  \caption{Distributions of reconstructed values for
    $\varepsilon_{\rm E}=\varepsilon_{\rm p}=5\%$ and
    $E_{\rm beam}=100~\text{GeV}$.}
  \label{fig:reconstructed_values}
\end{figure}

To illustrate this let us consider the following simple example. For the
monochromatic muon beam with energy $E_{\rm beam}$ two independent systems
measure the muons energy (analogous to forward detectors providing us with the
protons energy losses and therefore with particle energies) and momentum
(analogous to the central detector). Due to final resolutions of these systems,
$\varepsilon_{\rm E}$ and $\varepsilon_{\rm p}$, reconstructed energy $E_{\rm
rec}$ and momentum $p_{\rm rec}$ are not $\delta$-functions but distributed
around central values $E_{\rm beam}$ and $p_{\rm beam}=\sqrt{E_{\rm
beam}^{2}-m^{2}}$, where $m\approx 0.105~\text{GeV}$. For $\varepsilon_{\rm
E}=\varepsilon_{\rm p}=5\%$ the distributions for $E_{\rm rec}$ and $p_{\rm
rec}$ are shown in Fig.~\ref{fig:E_rec},~\ref{fig:p_rec}.

With measured energy and momentum for each muon, the mass can be reconstructed
as $m_{\rm rec}^{2}=E_{\rm rec}^{2}-p_{\rm rec}^{2}$. The corresponding
distribution is shown in Fig.~\ref{fig:m2_rec}. It has a peak at the squared
muon mass (indistinguishable from 0) as expected.

To plot \emph{the same data} with respect to $m_{\rm rec}=\sqrt{m_{\rm
rec}^{2}}$ one has to exclude events with negative $m_{\rm rec}^{2}$ and
calculate the square root for the remaining events. The result is shown in
Fig.~\ref{fig:m_rec}. It is clear that the maximum is shifted from 0. The reason
for that is just the change of variable for which the distribution was made. To
make it more evident we made the $m_{\rm rec}$ distribution with the larger bin
width and gave each bin its own color, see Fig.~\ref{fig:m_rec_highlight}. Then
we marked the events from these bins in the $m_{\rm rec}^{2}$ distribution by
the same color, see Fig.~\ref{fig:m2_rec_highlight}. Now it is clear why the
first bin in Fig.~\ref{fig:m_rec_highlight} is so small: in the $m_{\rm
rec}^{2}$ distribution it maps to a really narrow region. The next bin 10-20~GeV
maps to 100-400~GeV$^{2}$, i.e. three times wider than 0-100~GeV$^{2}$ for the
first bin 0-10~GeV.

If $p(m^{2})$ and $p'(m)$ are continuous probability distributions for $m_{\rm
rec}^{2}$ and $m_{\rm rec}$ value respectively, then we have the following
relation:
\begin{equation}
  \label{eq:m2_to_m}
  p'\left(m\right) = 2m\cdot p\left(m^{2}\right).
\end{equation}
Therefore, if $p\left(m^{2}\right)$ is regular at 0, then $p'\left(0\right)=0$.
This is just what can be seen in Figs~\ref{fig:m2_rec} and \ref{fig:m_rec}.

\begin{figure}[p]
  \centering
  \begin{subfigure}[b]{0.493\textwidth}
    \centering
    \includegraphics[width=3.3in]{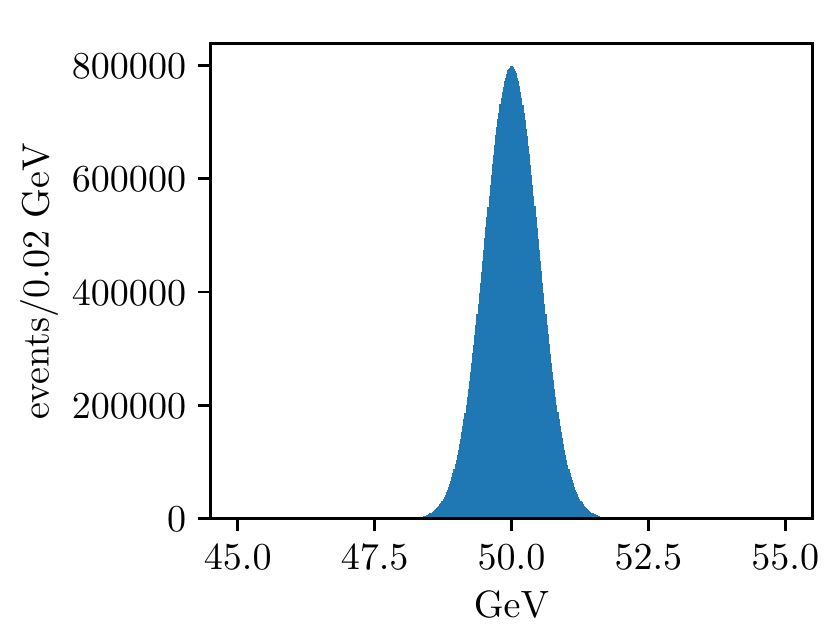}
    \caption{$E_{\rm rec}$.}
    \label{fig:E_rec2}
  \end{subfigure}
  \begin{subfigure}[b]{0.493\textwidth}
    \centering
    \includegraphics[width=3.3in]{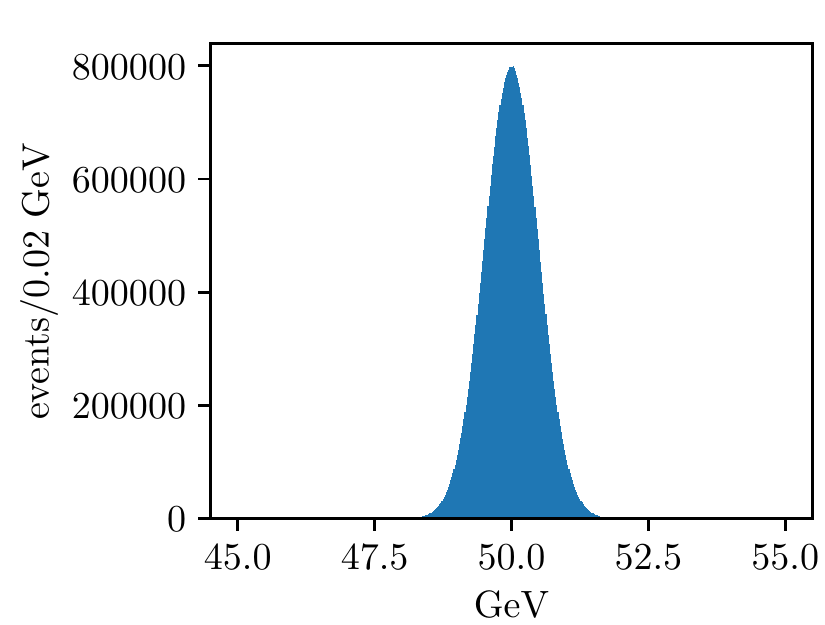}
    \caption{$p_{\rm rec}$.}
    \label{fig:p_rec2}
  \end{subfigure}\\
  \begin{subfigure}[b]{0.493\textwidth}
    \centering
    \includegraphics[width=3.3in]{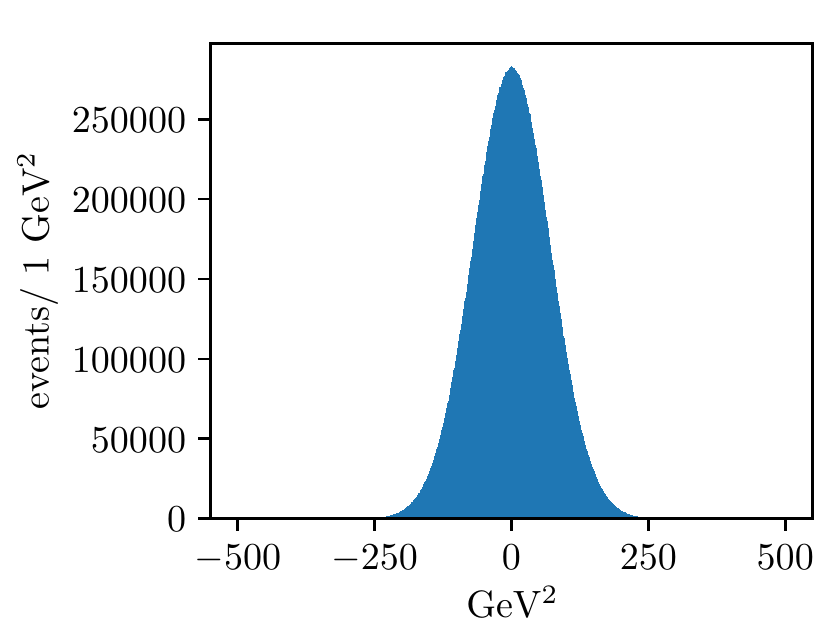}
    \caption{$m^{2}_{\rm rec}$.}
    \label{fig:m2_rec2}
  \end{subfigure}
  \begin{subfigure}[b]{0.493\textwidth}
    \centering
    \includegraphics[width=3.3in]{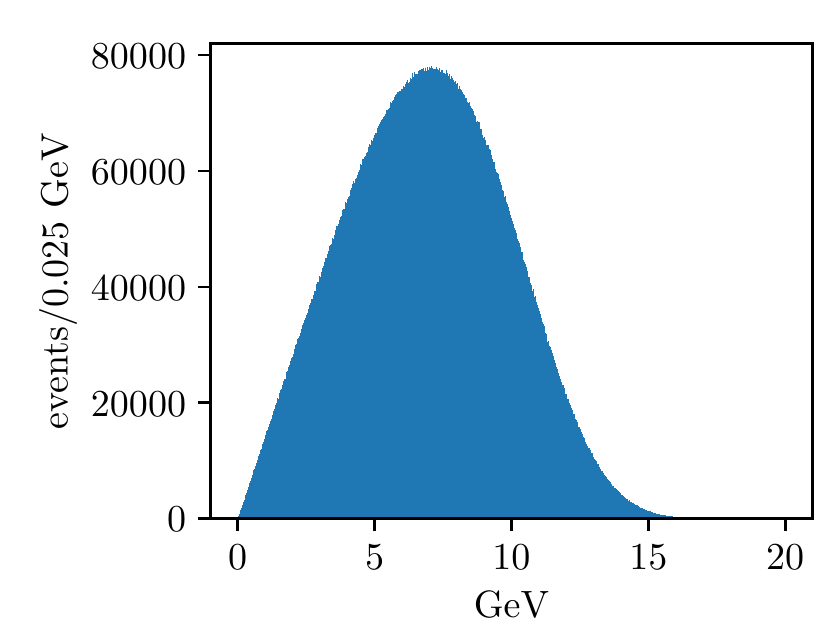}
    \caption{$m_{\rm rec}$.}
    \label{fig:m_rec2}
  \end{subfigure}\\
  \begin{subfigure}[b]{0.493\textwidth}
    \centering
    \includegraphics[width=3.3in]{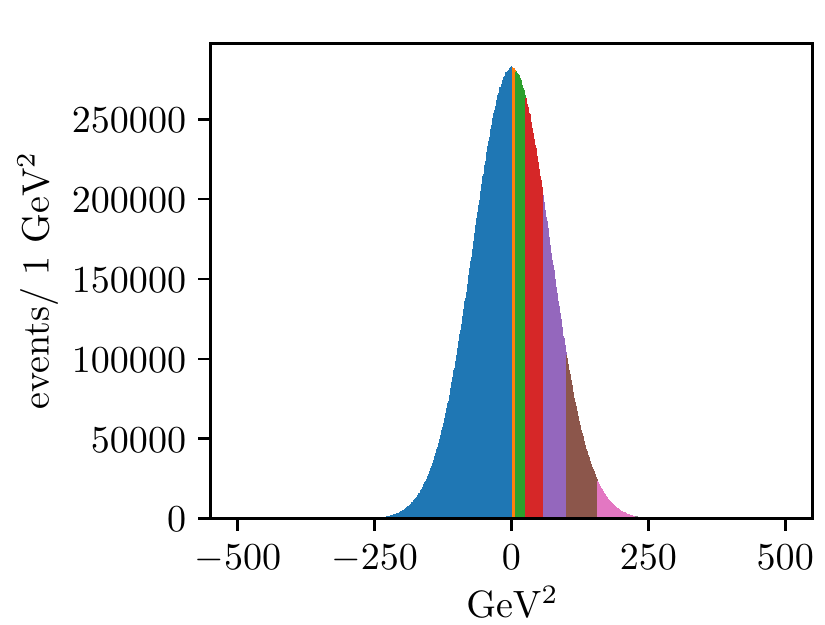}
    \caption{Highlighted $m^{2}_{\rm rec}$.}
    \label{fig:m2_rec_highlight2}
  \end{subfigure}
  \begin{subfigure}[b]{0.493\textwidth}
    \centering
    \includegraphics[width=3.3in]{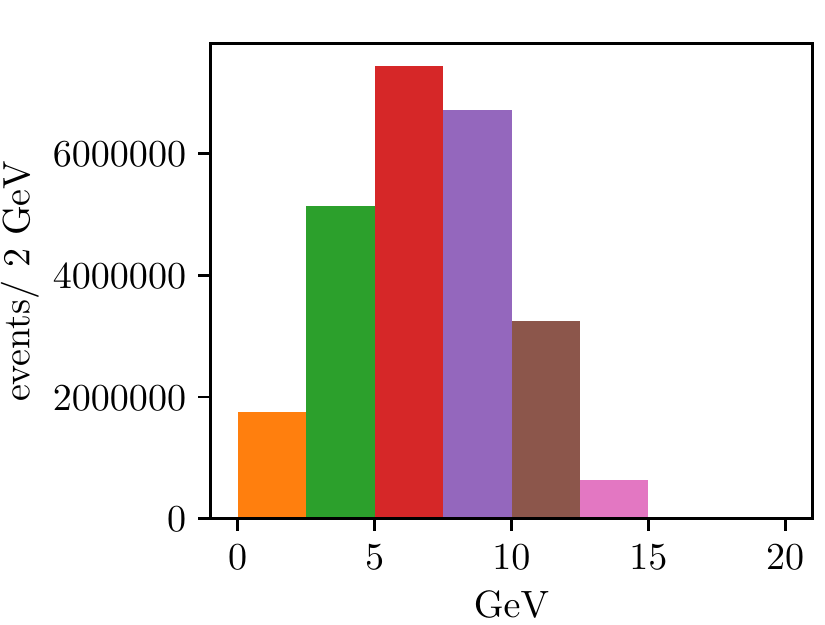}
    \caption{Highlighted $m_{\rm rec}$.}
    \label{fig:m_rec_highlight2}
  \end{subfigure}
  \caption{Distributions of reconstructed values for
    $\varepsilon_{\rm E}=\varepsilon_{\rm p}=1\%$ and
    $E_{\rm beam}=50~\text{GeV}$.}
  \label{fig:reconstructed_values2}
\end{figure}

The maximum of the $m_{\rm rec}$ distribution is defined by the width of the
$m_{\rm rec}^{2}$ distribution that originates from $E_{\rm rec}$ and $p_{\rm
rec}$ widths and central values.  To illustrate that we made the similar plots
for $\varepsilon_{\rm E}=\varepsilon_{\rm p}=1\%$ and $E_{\rm
beam}=50~\text{GeV}$, see Fig.~\ref{fig:reconstructed_values2}. The maximum has
shifted down to $7.5$~GeV.

\newcommand{\arxiv}[1]{\href{http://arxiv.org/abs/#1}{arXiv:\nolinebreak[3]#1}}

\end{document}